\documentclass[a4paper]{article}
\usepackage{bbold}
\usepackage[pages=all, color=black, position={current page.south}, placement=bottom, scale=1, opacity=1, vshift=5mm]{background}
\SetBgContents{
 
}      % copyright

\usepackage[margin=1in]{geometry} % full-width

\usepackage{amsmath}
\usepackage{amsthm}
\usepackage{amssymb}
\usepackage{multirow}
\usepackage{booktabs}

\usepackage[utf8]{inputenc}
\usepackage{hyperref}
\hypersetup{
	unicode,
}

\usepackage[sort&compress,numbers,square]{natbib}
\theoremstyle{plain}
\newtheorem{theorem}{Theorem}

\theoremstyle{definition}
\newtheorem{definition}[theorem]{Definition}

\usepackage{mathtools}% nccmath}

\usepackage{graphicx, color}
\graphicspath{{fig/}}

\usepackage{algorithm, algpseudocode} % use algorithm and algorithmicx for typesetting algorithms
\usepackage{mathrsfs} % for \mathscr command

\usepackage{lipsum}

\usepackage{color}
\title{Generalized contact matrices for epidemic modeling}

\author{Adriana Manna$^{1}$, Lorenzo Dall'Amico$^2$, Michele Tizzoni$^3$, M\'arton Karsai$^{1,4}$, Nicola Perra$^5$}

\date{
	$^1$ Department of Network and Data Science, Central European University, Vienna, Austria\\
	$^2$ ISI Foundation, Turin, Italy\\
	$^3$ Department of Sociology and Social Research, University of Trento, Trento, Italy\\
	$^4$ R\'enyi Institute of Mathematics, Budapest, Hungary\\ 
	$^5$ School of Mathematical Sciences, Queen Mary University of London, UK\\
	\vspace{0.5cm}
	\today
}

\begin{document}
	\maketitle

\begin{abstract}
Contact matrices have become a key ingredient of modern epidemic models. 
They account for the stratification of contacts for the age of individuals and, in some cases, the context of their interactions. However, age and context are not the only factors shaping contact structures and affecting the spreading of infectious diseases. Socio-economic status (SES) variables such as wealth, ethnicity, and education play a major role as well. Here, we introduce generalized contact matrices capable of stratifying contacts across any number of dimensions including any SES variable. We derive an analytical expression for the basic reproductive number of an infectious disease unfolding on a population characterized by such generalized contact matrices. 
Our results, on both synthetic and real data, show that disregarding higher levels of stratification might lead to the under-estimation of the reproductive number and to a mis-estimation of the global epidemic dynamics. 
Furthermore, including generalized contact matrices allows for more expressive epidemic models able to capture heterogeneities in behaviours such as different levels of adoption of non-pharmaceutical interventions across different groups. Overall, our work contributes to the literature attempting to bring socio-economic, as well as other dimensions, to the forefront of epidemic modeling. Tackling this issue is crucial for developing more precise descriptions of epidemics, and thus to design better strategies to contain them.
\end{abstract}

\maketitle

Contact matrices have become an integral part of realistic epidemic models for respiratory infections. Usually, they encode contact patterns through age-specific contact rates describing how frequently and for how long individuals of different ages meet each other. The stratification of contacts by age groups allows capturing heterogeneous mixing rates that are observed among individuals of different ages ~\cite{edmunds1997mixes,mossong2008social,prem2017projecting,rohani2010contact, Mistry2021,wallinga2006using}. Young adults, for example, are usually very active and tend to interact more with other young adults~\cite{hoang2019systematic}. Elderly individuals, instead, tend to report the fewest number of contacts and their role in the transmission dynamics of many respiratory infections is less relevant~\cite{wallinga2006using}. Contact matrices can be further stratified by the setting where interactions take place since the number and type of contacts vary considerably by context too (i.e. in household, workplace, school, community)~\cite{fumanelli2012inferring, Mistry2021}.

Mixing patterns disaggregated by age and context are key to estimating age-specific and context-specific transmission parameters for epidemic models, which are then used to guide health policy, evaluate intervention strategies and assess the risk of infection across population groups~\cite{hens2009estimating, goeyvaerts2010estimating, goeyvaerts2018household, aleta2022quantifying}. 
Despite their essential role, however, age and context are far from being the only important variables shaping contact patterns, disease dynamics, and epidemic outcomes. Individual characteristics related to socio-economic status (SES) such as wealth, race, ethnicity, occupation, and education, among others, have been shown to play a key role in the transmission of infectious diseases~\cite{buckee2021thinking,tizzoni2022addressing,bedson2021review,zelner2022there}. From the Influenza pandemics of 1918 and 2009~\cite{grantz2016disparities, mamelund2021association} to the West African Ebola outbreak~\cite{alexander2015factors} and the COVID-19 pandemic~\cite{perra2021non,garnier2021socioeconomic,zelner2022there,do2021association,gozzi2021estimating,heroy2021covid,gauvin2021socio,weill2020social,jay2020neighbourhood,bonaccorsi2020economic,fox2023disproportionate,gozzi2023estimating} belonging to lower SES has been consistently associated with higher rates of infections, deaths, as well as reduced access to care and ability to comply with non-pharmaceutical interventions (NPIs).

Despite the recognised importance of SES in the transmission dynamics of close-contact infections, the overwhelming majority of epidemic models neglect these dimensions~\cite{buckee2021thinking,tizzoni2022addressing,bedson2021review,zelner2022there}. Socio-economic status is often considered only \emph{a posteriori} in analyses of models' outputs (e.g., number of deaths or cases) but rarely enters at the core of their formulation as age does. The roots of this shortcoming can be traced back to the lack of i) modelling frameworks designed to consider SES (or other variables) as one, or more, of their structural features~\cite{zelner2022there}, ii) empirical surveys of social contact data including characteristics of the respondents to account for SES~\cite{mousa2021social, hoang2019systematic}.  

Here, we tackle the first limitation by developing a general epidemic framework able to accommodate \emph{generalized contact matrices} stratified by multiple dimensions. 
We focus on prototypical Susceptible-Exposed-Infectious-Recovered (SEIR) compartmental models and derive a closed-form expression for the basic reproductive number, $R_0$, for any number of dimensions. In doing so, we quantify how much the $R_0$ computed from a generalized contact matrix differs from classic models considering only one demographic attribute such as age. By inspecting the spectral properties of the classic and generalized contact matrices, we prove that models fed with the latter are characterized by a $R_0$ which is necessarily either equal to or higher than the corresponding value computed in single-attribute models. Interestingly, for a given set of disease parameters, we find that correlations in contact patterns, such as mixing assortativity and heterogeneous activity, increase the value of $R_0$. After validating our mathematical formulation with numerical simulations in theoretical scenarios, we showcase how the use of generalized contact matrices allows for more expressive epidemic models able to describe the effects of NPIs during an outbreak accounting for possible heterogeneities in their adoption across population subgroups. 
Finally, we apply our approach to real data from Hungary and we quantify the possible misrepresentation induced by neglecting SES in simulated outbreaks. The results confirm significant differences between models. The use of generalized contact matrices, that stratify contacts by age and one SES dimension, in accordance with the mathematical derivations, leads to higher values of $R_0$ for a given disease, to lower values of attack rates for a given $R_0$, and allows to capture heterogeneity in disease's burden across SES groups.

\section*{Results}

We consider a Susceptible-Exposed-Infectious-Recovered (SEIR) compartmental model where susceptible ($S$) are healthy individuals at risk of infection, exposed ($E$) are infected but not yet infectious, infectious ($I$) are able to spread the disease, and recovered ($R$) are no longer infectious nor susceptible to the disease~\cite{keeling08}. Within this setting, we propose a modelling framework that extends classic approaches by including generalized contact matrices that stratify contacts across multiple dimensions. We provide an overview of standard contact matrices in epidemic models in the Supplementary Information.

\begin{figure*}
    \centering
    \includegraphics[scale=0.24]{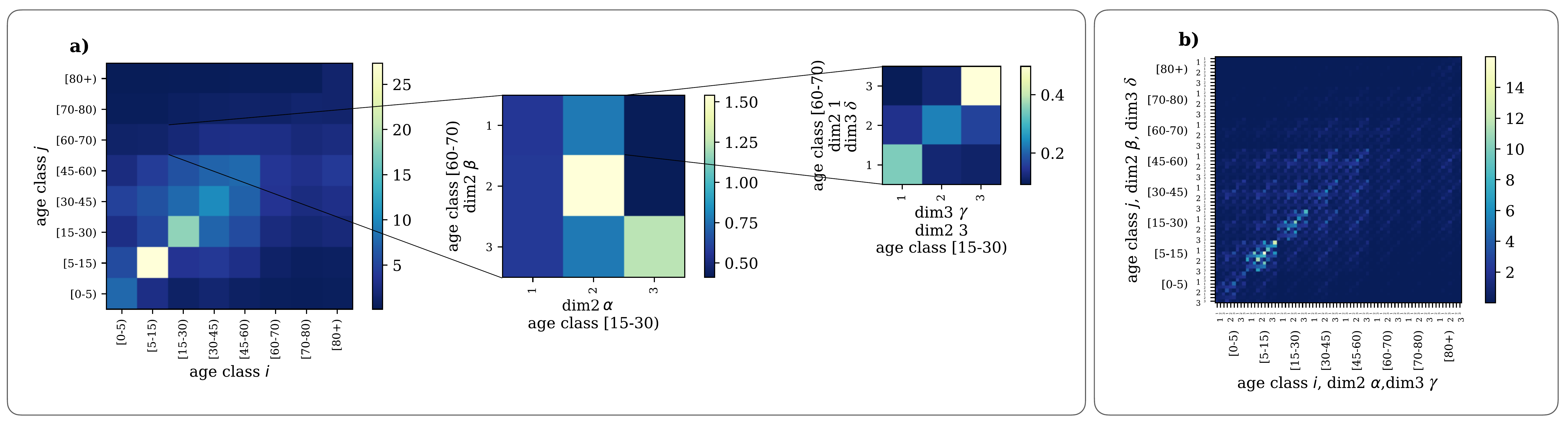}
    \caption{\textbf{Generalized contact matrices}. a) Schematic representation of generalized contact matrices. We can transform a $K \times K$ age-structured contact matrix $\mathbf{C}$ (first matrix on the left) into a generalized matrix $\mathbf{G}$ with $m$ new dimensions. Such transformation can be done according to different generalization schemes that we discuss in the Methods and the Supplementary Information. In the plot, we consider $m=2$. Hence, the second matrix describes the stratification of contacts across a second dimension i.e., $\alpha, \beta \in [1,2,3]$. The third matrix describes a further stratification, i.e., $\gamma, \delta \in [1,2,3]$. For simplicity we show  the stratification across the new dimensions only for one entry of the lower level respectively.
    b) Example of a flatten generalized contact matrix including $3$ dimensions $\mathbf{G}$ ($72 \times 72$).} 
    \label{panel_1}
\end{figure*}

\subsection*{Generalized contact matrices} 
We present an extension of classic epidemic models by introducing generalized contact matrices further stratified by other dimensions besides age or context. In what follows, we consider indicators of SES as new dimensions, but the formulation is general and could accommodate any other relevant categorical feature of the population under study. 
The proposed framework goes from the classic contact matrices $C_{ij}$, where individuals are grouped according to their age bracket $(i, j)$, to $G_{\mathbf{a},\mathbf{b}}$, where individuals are characterized by their age and $m$ other categorical variables. 
Hence, $\mathbf{a}=(i,\alpha, \beta,\ldots,\gamma)$ and $\mathbf{b}=(j,\eta, \nu,\ldots,\xi)$ are index vectors (i.e., tuples) representing individual's membership to each category. We use Greek letters to highlight the difference between age and other dimensions. 
To provide a concrete example of a generalized contact matrix, we can imagine stratifying the population according to age ($i \in [1,\ldots, K]$), income ($\alpha \in [1,\ldots, V_1]$), and education attainment ($\beta \in [1,\ldots, V_2]$). 
In this case the generalized contact matrix $G_{\mathbf{a},\mathbf{b}}$ would describe the average number of contacts that an individual in age bracket $i$, income $\alpha$, and education $\beta$ has with individuals in age group $j$, income $\eta$, and education $\nu$ in a given time window (see Figure~\ref{panel_1}). 
From this perspective, the matrix $C_{ij}$ can be thought of as an aggregation of contact rates at lower levels of stratification. 
As with any aggregated metric, the elements $C_{ij}$ are agnostic about the structure of contacts across other dimensions. 

We denote with $K$ the number of age groups, while with $V_p$ the $p\in[1,\ldots, m]$ the number of groups in each of the other dimensions. While the generalized matrix $\mathbf{G}$ (whose elements are $G_{\mathbf{a},\mathbf{b}}$) can be naturally described as a multi-dimensional object, the use of $T=K\prod_{p=1}^{m}V_p \times K\prod_{p=1}^{m}V_p$ index vectors allows for a flattened representation in a squared 2-dimensional (2D) matrix of size $\sqrt{T} \times \sqrt{T}$. 
We note how the formulation can easily consider different contexts where interactions take place, i.e., $G_{\mathbf{a},\mathbf{b}}=\sum_{l}\omega_l G^{(l)}_{\mathbf{a},\mathbf{b}}$, where $\omega_l$ captures the relative importance of the different social settings in the transmission~\cite{Mistry2021}.

\subsection*{Epidemic model} 
As mentioned above, we study diseases whose natural history can be described by an SEIR compartmental model. Similar results can be obtained for others such as the SIS and SIR models. To include the generalized contact matrices $\mathbf{G}$, we define $X_{\mathbf{a}}$ as the number of individuals in groups $\mathbf{a}$ and compartment $X$ where $X\in [S,E,I,R]$. The dynamics of the model are described by the following set of differential equations:
\begin{eqnarray}
d_t S_{\mathbf{a}}(t)&=&-\Lambda_{\mathbf{a}}(t)S_{\mathbf{a}}(t) \nonumber \\
d_t E_{\mathbf{a}}(t)&=&\Lambda_{\mathbf{a}}(t)S_{\mathbf{a}}(t)-\Psi E_{\mathbf{a}}(t) \nonumber \\
d_t I_{\mathbf{a}}(t)&=& \Psi E_{\mathbf{a}}(t)-\Gamma I_{\mathbf{a}}(t) \nonumber\\
d_t R_{\mathbf{a}}(t)&=&\Gamma I_{\mathbf{a}}(t)
\end{eqnarray}
where $\Gamma$ is the recovery rate, $\Psi$ is the rate at which exposed become infectious, the force of infection $\Lambda_{\mathbf{a}}(t)=\Phi \sum_{\mathbf{b}} G_{\mathbf{a},\mathbf{b}} \frac{I_{\mathbf{b}}(t)}{N_{\mathbf{b}}}$ describes the per capita rate at which susceptible individuals in categories $\mathbf{a}$ acquire the infection, and $\Phi$ is the transmissibility of the disease. 
For simplicity of exposition, we neglect demography (i.e, birth and death rates) hence the size of the population is fixed. To avoid confusion with the indices of the additional dimensions, the parameters regulating the disease dynamics are denoted with capital Greek letters.

\subsection*{Basic reproductive number} The basic reproductive number $R_0$ is one of the most important quantities of epidemiological relevance~\cite{keeling08}. It is defined as the number of secondary infections generated by a single infected seed in an otherwise susceptible population. In the case of standard contact matrices $\mathbf{C}$, the $R_0$ of an SEIR model can be obtained using the next-generation matrix approach (see Refs.~\cite{diekmann2010construction,blackwood2018introduction} for overviews of this method). The expression reads $R_0=\frac{\Phi}{\Gamma}\rho(\tilde{\mathbf{C}})$, where $\rho$ denotes the spectral radius of the matrix $\tilde{\mathbf{C}}$ whose elements are $\tilde{C}_{ij}=\frac{C_{ij}}{N_j}N_i$, $N_i$ and $N_j$ describe the number of individuals in age-bracket $i$ and $j$ respectively. How does this expression change when we consider generalized contact matrices? In the following, we show that the generalisation to contact matrices featuring any number of dimensions can be still tackled with the next-generation matrix approach. First, we focus our attention only on the compartments $E$ and $I$, which capture individuals in one of the two possible stages of infection. Furthermore, as a first step, we consider the simplest form of generalized contact matrices featuring only one additional variable besides age, which we generally denote as an SES variable, i.e., $\mathbf{a}=(i,\alpha)$. 
Let us define $\mathbf{x}$ as a vector of size $2KV_1$ such that $\mathbf{x}=(E_{11},E_{12},\ldots E_{KV_1},I_{11},I_{12},\ldots I_{KV_1})$, where, for simplicity, we omitted the time dependence. 
We can express the set of differential equations for these compartments as $d_t \mathbf{x}=f(\mathbf{x})-w(\mathbf{x})$, where $f(\mathbf{x})$ describes all terms that lead to new infections and $w(\mathbf{x})$ all other transitions in and out of the compartments. The expression of $R_0$ is linked to the early epidemic dynamics, which can be linearized by calculating the Jacobian at the disease-free equilibrium (DFE) $(S^*_{i\alpha},E^*_{i\alpha},I^*_{i\alpha},R^*_{i\alpha})=(N_{i\alpha},0,0,0)$ for all age groups and SES. 
Note how $N_{i\alpha}$ is the population in age-group $i$ and SES $\alpha$. Using the Jacobian matrix $\mathbf{J}$ (which has size $2KV_1\times 2KV_1$) at the DFE we can write $d_t \mathbf{x}=\mathbf{J}\mathbf{x}$. This expression can be conveniently factorized as $d_t \mathbf{x}=(\mathbf{F}-\mathbf{W})\mathbf{x}$ where the matrix $\mathbf{F}$ is Jacobian applied to $f(\mathbf{x})$ and similarly the matrix $\mathbf{W}$ is the Jacobian applied to $w(\mathbf{x})$. As shown in the Supplementary Information, the two matrices can be expressed as:
\begin{equation}
    \mathbf{F} = \Phi
\begin{pmatrix}
\mathbf{0} & \mathbf{\tilde{G}}\\
\mathbf{0} &\mathbf{0} 
\end{pmatrix}, \;\ \mathbf{W} = 
\begin{pmatrix}
\Psi \mathbf{I} & \mathbf{0}\\
-\Psi \mathbf{I} &\Gamma \mathbf{I}
\end{pmatrix}
\end{equation}
where $\mathbf{0}$ describes square blocks of zeros, $\mathbf{I}$ is the identity matrix, and $\tilde{\mathbf{G}}$ is a matrix whose element are $\tilde{G}_{i \alpha, j\beta}=\frac{G_{i \alpha, j\beta}}{N_{j\beta}}N_{i \alpha}$. All these matrices have a size of $KV_1\times KV_1$. Following the next-generation approach, we can obtain a closed-form expression for the basic reproductive number:
\begin{equation}
\label{R0}
  R_0=\rho(\mathbf{FW}^{-1})=\frac{\Phi}{\Gamma}\rho(\mathbf{\tilde{G}})
\end{equation}
It is interesting to note how the structural features of all matrices, and thus $R_0$, are analogous to the simple case of $\mathbf{C}$. The key difference is in the size and composition of the matrices. Indeed, we shift from indices $i$ and $j$ to tuples $\mathbf{a}=(i,\alpha)$ and $\mathbf{b}=(j,\beta)$ describing the contacts across groups and dimensions. As shown in the Supplementary Information, the same structure holds for any size of the index vectors. Hence, Eq.~\ref{R0} is valid for any number of dimensions where $\tilde{G}_{\mathbf{a}, \mathbf{b}}=\frac{G_{\mathbf{a},\mathbf{b}}}{N_{\mathbf{b}}}N_{\mathbf{a}}$, $\mathbf{a}=(i,\alpha,\ldots,\gamma)$, and $\mathbf{b}=(j,\eta,\ldots,\xi)$. Eq.~\ref{R0} is valid also in case contacts are further stratified according to the context where they take place (see Supplementary Information). 
As mentioned above, the contact matrix $\mathbf{C}$ can be considered as an aggregation of its generalized counterpart $\mathbf{G}$, along all dimensions but the age. Since the total number of contacts in the population is independent of the aggregation, we can write $C_{ij}N_i=\sum_{\mathbf{a' b'}}G_{\mathbf{a'},\mathbf{b'}}N_{\mathbf{a'}}$, where $\mathbf{a'}=\mathbf{a}-\{i\}$ and $\mathbf{b'}=\mathbf{b}-\{j\}$ are the index vectors of size $m$ capturing all dimensions but age. Similarly, we can obtain an expression for the average number of contacts that individuals in one SES, say $\alpha$, have with others in the same SES $\beta$ aggregating over all other groups $C_{\alpha \beta}N_\alpha=\sum_{\mathbf{a' b'}}G_{\mathbf{a'},\mathbf{b'}}N_{\mathbf{a'}}$, where now $\mathbf{a'}=\mathbf{a}-\{\alpha\}$ and $\mathbf{b'}=\mathbf{b}-\{\beta\}$. In general, $C_{ij}\neq C_{\alpha \beta}$. 
Indeed, the number of groups and the number of individuals in each of them might be different. Looking at contact patterns from multiple dimensions of stratification highlights how the way interactions are aggregated (i.e., by age or other dimensions) might affect the estimation of key epidemiological parameters and the epidemic dynamics. 
For example, the spectral radius of standard contact matrices, as we saw, directly influences the basic reproductive number of the disease. Interestingly, in the Supporting Information, we prove how the spectral radius of generalized contact matrices is always larger than or equal to the corresponding value of the traditional and aggregated contact matrices. Hence, the estimation of $R_0$ might be affected by the dimensions that are aggregated and also by those, which are explicitly considered in the stratification. In particular, we show that strict equality can be attained under the random mixing hypothesis, i.e., when the contacts are set proportional to the product of the subpopulation sizes. Overall, the variables used to capture the stratification of contacts might affect the description of epidemic processes, especially in the presence of non trivial correlations between explicit and implicit (i.e., aggregated) variables. Indeed, the estimation of the spreading potential of disease via a model is not only a function of the total number of contacts but also of how these contacts are arranged across groups. This observation is one of the key insights from network-based epidemiology: the epidemic threshold (i.e., $R_0$) in two populations with the same number of individuals and number of contacts is drastically influenced by the way interactions are organized (i.e., the topology of the network)~\cite{barrat2008dynamical}.

\begin{figure*}
    \centering
    \includegraphics[scale=0.42]{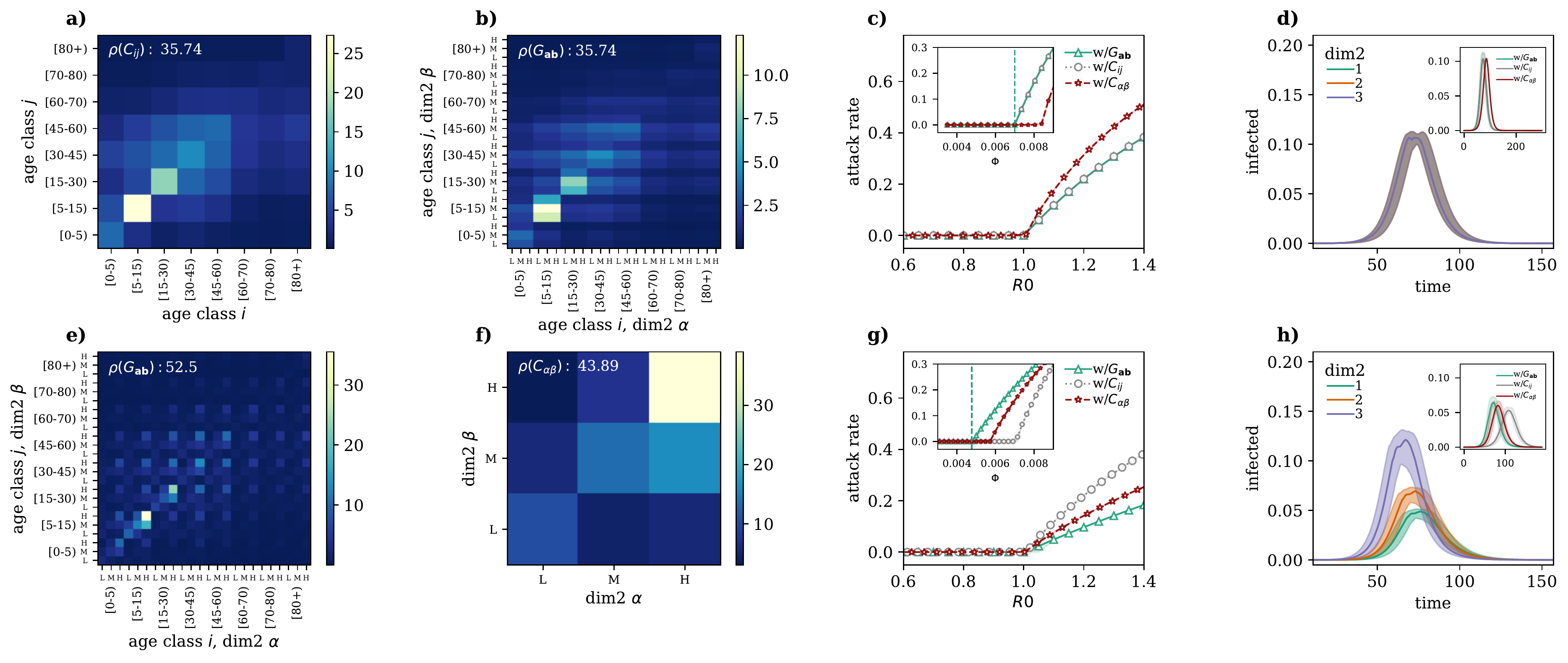}
    \caption{\textbf{Impact of generalized contact matrices on epidemic spreading}. (a) Example of a classic age contact matrix $\mathbf{C}$ ($8 \times 8$); (b) and (e) generalized contact matrices with an additional dimension with three groups $\mathbf{G}$ ($24 \times 24$) in the case of random mixing (b) and in the case of assortative mixing (e) in the second dimension; (f) depicts the case when we integrate the generalized contact matrix over all age classes. The $\rho$ values indicate the spectral radius corresponding to each matrix. In panel (c) we show the attack rate as a function of $R_0$ and disease prevalence. In panel (d) the attack rate is shown as a function of time in the case of random mixing; (g) and (h) are as the previous two plots but for the case of assortative mixing with different levels of activity. Results refer to the median of $500$ runs. Epidemiological parameters: $\Gamma=0.25, \Psi=0.4$, and $R_0=2.7$. Simulations start with a number $I_0= 100$ of initial infectious seeds.}  \label{panel_2}
\end{figure*}

\subsection*{Numerical simulations: synthetic data} We test the analytical results obtained in the previous section by considering synthetic generalized contact matrices first. To ground the results with some empirical observations, we build on real traditional contact matrices, where interactions are stratified only according to age groups. Here, we rely on data from Hungary by using a pre-COVID-19 age contact matrix~\cite{koltai2022reconstructing}, while we also report analogous findings for a different country in the Supplementary Information. We create generalized contact matrices by adding more dimensions to the empirical matrices and by defining the contact rates in the added groups with a simple model. We explore two cases. In the first case, contact rates for the additional dimensions are set to be proportional to the product of the population sizes of the different groups. This is the aforementioned random mixing. In the second case, instead, we investigate scenarios where contact rates among subgroups are defined by parameters. This allows us to introduce correlations, such as assortativity, where members in a given group (e.g., in a SES class) are more likely to contact people in the same group. Furthermore, in the second case, we introduce activity parameters to regulate the activity (i.e., share of contacts) of the different groups. As shown in Fig.~\ref{panel_2}, when considering real data from Hungary with an additional dimension, non trivial correlations and heterogeneity appear. We refer the reader to the Material and Methods and the Supplementary Information for details about the construction of the synthetic generalized contact matrices. 
In Fig.~\ref{panel_2}-a we display the real age contact matrix $\mathbf{C}$ for Hungary, which stratifies interactions in $K=8$ age brackets. While until the 45-60 age group, the highest values of contact rates are within the same age bracket (i.e., diagonal elements), we observe also strong off-diagonal values for age groups that range from 15 to 60 years. In Fig.~\ref{panel_2}-b we show the flatten 2D representation of a generalized contact matrix where, besides age, we have a second dimension. We imagine a simple case where the second dimension contains three groups, i.e., $V_1=3$. We assume that respectively $35\%$, $45\%$, and $20\%$ of the population belong to these three categories across all age groups for simplicity. The matrix is formed by $K\times V_1=24$ distinct groups. 

\subsubsection*{Random mixing scenario} First,  we consider that contacts -- when it comes to the second dimension -- are proportional to the product of the population size in each group. Hence, they are the expected value of a random mixing process. As mentioned above, this assumption leads to a generalized contact matrix with the same spectral radius as the original contact matrix. The effects of this property on an epidemic model, that used to study the unfolding of a virus in the population, are shown in Fig.~\ref{panel_2}-c. We plot the attack rate (i.e., epidemic size) as a function of $R_0$ for i) a model fed only with the contact matrix $C_{ij}$ (grey circles), ii) a model fed with the generalized contact matrix $G_{\mathbf{a},\mathbf{b}}$ described above (green triangles), iii) a model fed with the matrix $C_{\alpha \beta}$ where contacts are stratified only according to the second dimension (red stars). As a result we observe, in perfect agreement with the analytical formulation, that $R_0=1$ divides the phase space into two regions. For values below this threshold, the disease is not able to spread in any of the three models. Only for values larger than one, the disease affects a finite fraction of the population. We observe that the threshold and the attack rates for different values of $R_0$ are the same in the first two models. Hence, the description of the epidemic in a model that either neglects or considers a second dimension of contact stratification does not change in case of random mixing in the second dimension of contacts. As a consequence, the evolution of the fraction of infected individuals in the groups of the additional category is also the same (see Fig.~\ref{panel_2}-d). We note however clear differences with respect to the third model (red stars) that features a $3\times 3$ contact matrix capturing the stratification of contacts only for the additional category. A description of the epidemic in these settings leads to higher attack rates for any given $R_0>1$. This observation confirms how within the same population the chosen stratification of contacts might affect the description of an epidemic unfolding in the system. In the inset of Fig.~\ref{panel_2}-c we show the attack rate as a function of the transmissibility $\Phi$. The vertical dashed line denotes the analytical critical value of $\Phi$ obtained by setting Eq.~\ref{R0} equal to one for the model featuring the generalized contact matrix. The plot shows the equivalence between the first model and the second as well as the difference with respect to the third.

%\medskip

\subsubsection*{Assortative scenario} We now consider the second scenario: assortative mixing. In Fig.~\ref{panel_2}-e we display the 2D flattened generalized contact matrix obtained by assuming that $60\%$, $50\%$, and $65\%$ of the contacts in the first, second, and third group of the additional category take place within each group, this way inducing assortativity. Furthermore, we assume some levels of heterogeneity also in the activity of the different groups setting as $20\%$, $40\%$, and $40\%$ the share of contacts of the three groups, respectively. Even a qualitative visual inspection of the generalized contact matrix shows how the introduction of assortativity and activity visibly changes the contact rates (see panel Fig.~\ref{panel_2}-e) with respect to the previous case of random mixing (panel Fig.~\ref{panel_2}-b). This is confirmed by the value of the spectral radius which is around $46\%$ larger, increasing from $\rho(G_{\mathbf{ab}})=35.74$ to $52.5$. In Fig.~\ref{panel_2}-f we show the matrix $C_{\alpha \beta}$ obtained after integrating the generalized contact matrix over all age classes. As imposed by construction, the third group is characterized by a higher assortativity than the others. It is the group featuring the smallest fraction of the population, and is the most active (together with the second group). These characteristics explain the high diagonal value in the matrix describing the contact rate between people in that group. As a result, the spectral radius is $24\%$ higher (being $\rho(C_{\alpha\beta})$=43.89) with respect to the contact matrix stratified for age (with $\rho(C_{ij})$=35.74). The impact of the different contact matrices on the estimated attack rate is shown in Fig.~\ref{panel_2}-g where we plot it as function of $R_0$. The figure shows in each case that the critical point falls at $R_0=1$, thus the analytical solutions match the numerical simulations very well.
%Indeed, also in this case the phase space is divided in two regions separated by the critical value $R_0=1$.
Contrary to the previous case of random mixing, the attack rate of the model that is informed with the contact matrix $C_{ij}$ and with the generalized matrix $G_{\mathbf{a},\mathbf{b}}$ are different for $R_0>1$. Interestingly, the strong assortativity of contacts in the population constrains the spreading of the disease, resulting in a smaller fraction of the population affected by the spreading (see the green triangles). Furthermore, the most active group is also the smallest one in this setting. A similar behaviour is observed on contagion processes unfolding on explicit contact networks organized in clusters, where the local topological correlations might slow down the spreading of contagion processes that in turn do not evolve into an endemic state (i.e., SIR-like dynamics)~\cite{sun2015contrasting}. In case the contacts are stratified only according to the second dimension (red stars), the attack rates are closer to those emerging from the generalized contact matrices (green triangles). In the inset of the figure we show the attack rate as function of the transmissibility $\Phi$. The vertical dashed line denotes the theoretical prediction of its critical value when considering the generalized contact matrices. Due to the differences in the spectral radius of the various matrices, the critical values of $\Phi$ do not coincide. In the setting considered here, neglecting the assortativity and activity across the second dimension in favour of simpler representations focused only on age or the additional variable leads to a possibly large underestimation of the critical value of $\Phi$. Assortativity and activity introduce variations also in the burden of the diseases across the secondary dimension. Indeed, Fig.~\ref{panel_2}-h shows how, in these settings, individuals in the third class are affected earlier and more intensely than the others. This is due to their high activity and assortativity. Finally, in the inset of Fig.~\ref{panel_2}-h we show the incidence for three models fixing a given value of $R_0=2.7$. Those standard contact matrices feature a smaller and later peak with respect to the other two. Even though these results are driven by the assumptions made when constructing the generalized contact matrix, they show how descriptions of epidemics with models, that include or neglect further stratification of contacts beside age, might be extremely different. In the Supplementary Information, we report a larger exploration of the parameter spaces, which confirms the validity of the analytical solutions. In addition, we include also scenarios with additional dimensions. The results confirm that Eq.~\ref{R0} provides a good description of the epidemic dynamics even in generalized contact matrices with three dimensions.

\begin{figure*}
    \centering
    \includegraphics[scale=0.70]{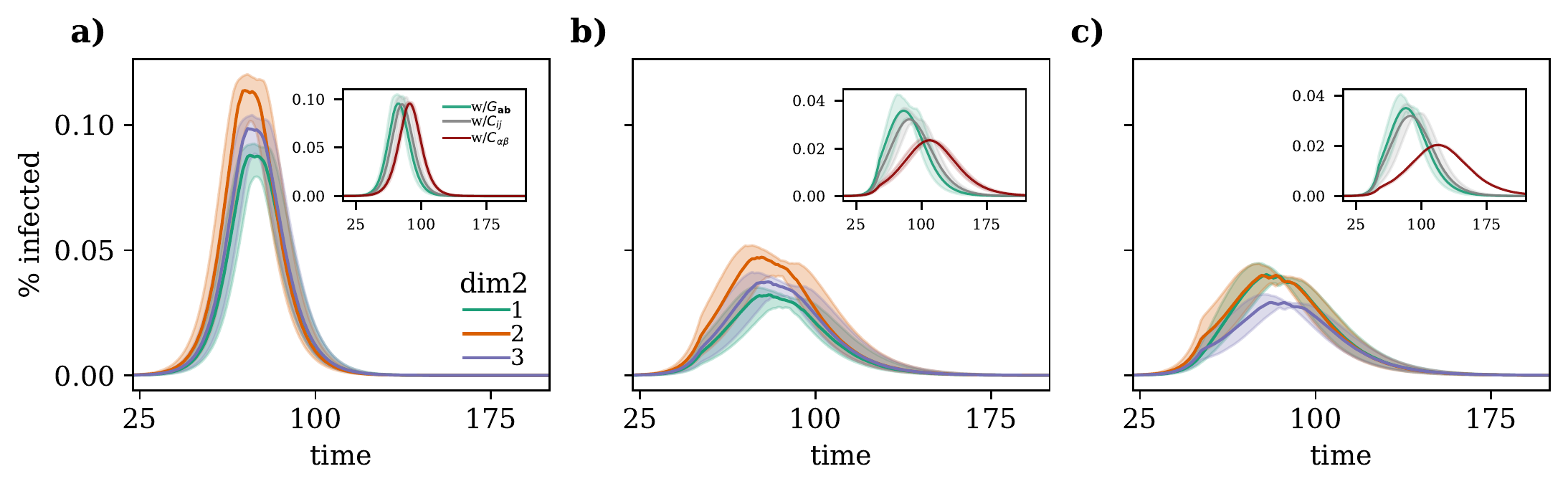}
     \caption{\textbf{Modeling the adoption of non-pharmaceutical interventions (NPIs)}. Disease prevalence in (a) baseline scenario (b) and (c) with NPIs introduced at $t^*=50$ that induce reduction of $35\%$ in total number of contacts with respect to the baseline scenario. In case of (b) the three groups reduce equally the number of contacts, while in (c) the NPIs induce a redistribution of the number of contacts. Curves in the main plot indicate the  prevalence for the three groups of the additional dimension, while in the inset they represent the prevalence for models featuring generalized contact matrices (green line), standard age-stratified contact matrices (gray line), and contact matrices stratified by the additional dimensions (red line). Results refer to the median of $500$ runs with confidence intervals. Epidemiological parameters: $\Gamma=0.25, \Psi=0.4$, and $R_0=2.7$. Simulations start with $I_0= 100$ initial infectious seeds.}
    \label{panel_3}
\end{figure*}

\subsubsection*{Modeling the adoption of non-pharmaceutical interventions} As mentioned above, epidemic models featuring generalized contact matrices are more expressive with respect to traditional approaches. They allow capturing possible heterogeneities in behaviors that might span from the adoption of NPIs, to vaccination uptake across diverse groups of the population. Indeed, the COVID-19 pandemic has been a stark reminder that the ability to comply with NPIs and vaccination rates are far from homogeneous. They correlate and are influenced by many SES variables~\cite{perra2021non,tizzoni2022addressing,buckee2021thinking,gozzi2023estimating}. Standard models allow describing such heterogeneities only partially, considering for example contact reductions as a function of age and location~\cite{di2020impact}. 

To showcase the potential of generalized contact matrices in this context, we explore hypothetical scenarios where a given population, featuring an additional dimension beside age, is changing behavior as a result of the introduction of NPIs. We consider the ability to reduce contacts to be related to the membership of a particular group of the population. We start with a pre-epidemic baseline where $33\%$, $50\%$ and $60\%$ of the contacts in the first, second, and third group of the additional category takes place within each group. In the baseline $25\%$, $45\%$, and $30\%$ are the shares of contacts of the three groups. For simplicity, we assume an even distribution of the population across the three groups. In Fig.~\ref{panel_3}-a, we show the incidence of a disease that unfolds undisturbed by any NPIs in a population described by such baseline. Individuals in the second group are the most affected as result of their higher activity. The inset shows the estimation of the overall incidence considering, as before, the three epidemic models featuring generalized contact matrices (green line), standard age-stratified matrices (grey line), and contact matrices stratified only for the second dimension (red line). In these settings the first model peaks a bit earlier, but the three curves are overall very similar.

Then, we imagine two scenarios where at time $t^*=50$, due to NPIs, the population changes behaviors. We refer to the Supporting Information for the details about the implementation. In Fig.~\ref{panel_3}-b we show what would happen to the epidemic in case of a homogeneous reduction scenario, when all groups would be able to reduce their contacts by $35\%$. The impact of the NPIs is quite strong and induces a clear reduction of the peak across the three groups of the population. The inset of Fig.~\ref{panel_3}-b shows how in this scenario the overall incidence, estimated with a model featuring generalized contact matrices (green line), peaks earlier and higher than with the other two. The model fed with standard age-stratified contact matrices (gray line) is very similar. The third model (red line) instead is rather different as the peak is not only lower but also delayed. This is another reminder of how the representation of contacts, the choice of which variables are considered or integrated, might affect the estimation of the epidemic dynamics.

In Fig.~\ref{panel_3}-c, we explore a scenario where the first group (e.g. representing the lowest SES class) is not able to protect itself as the other two manage to afford. We model this by reducing the overall number od contacts by $35\%$ but assuming that the NPIs introduce changes in both assortativity and activity. We imagine that due to the NPIs $50\%$, $60\%$ and $70\%$ of the contacts in the first, second, and third group of the additional category take place within each group. Hence, across the groups, assortativity increases, but more significantly for the second and third group. Furthermore, we imagine that the NPIs shift activities to $37\%$, $37\%$, and $26\%$ across the three groups. Hence, while activity decreases for the second and third group, it relatively increases for the first. Observing the incidence curves, we see how the NPIs bring a general reduction, but the changes in the contact patterns affect the first group more negatively. Indeed, we observe a switch from a baseline where this group was the least affected to a scenario where it becomes the most affected (together with the third group). The inset of the figure shows the overall incidence for the three models which is very close to the previous.

\subsection*{Numerical simulations: real data} 
\begin{figure*}
    \centering
    \includegraphics[width=0.83\textwidth]{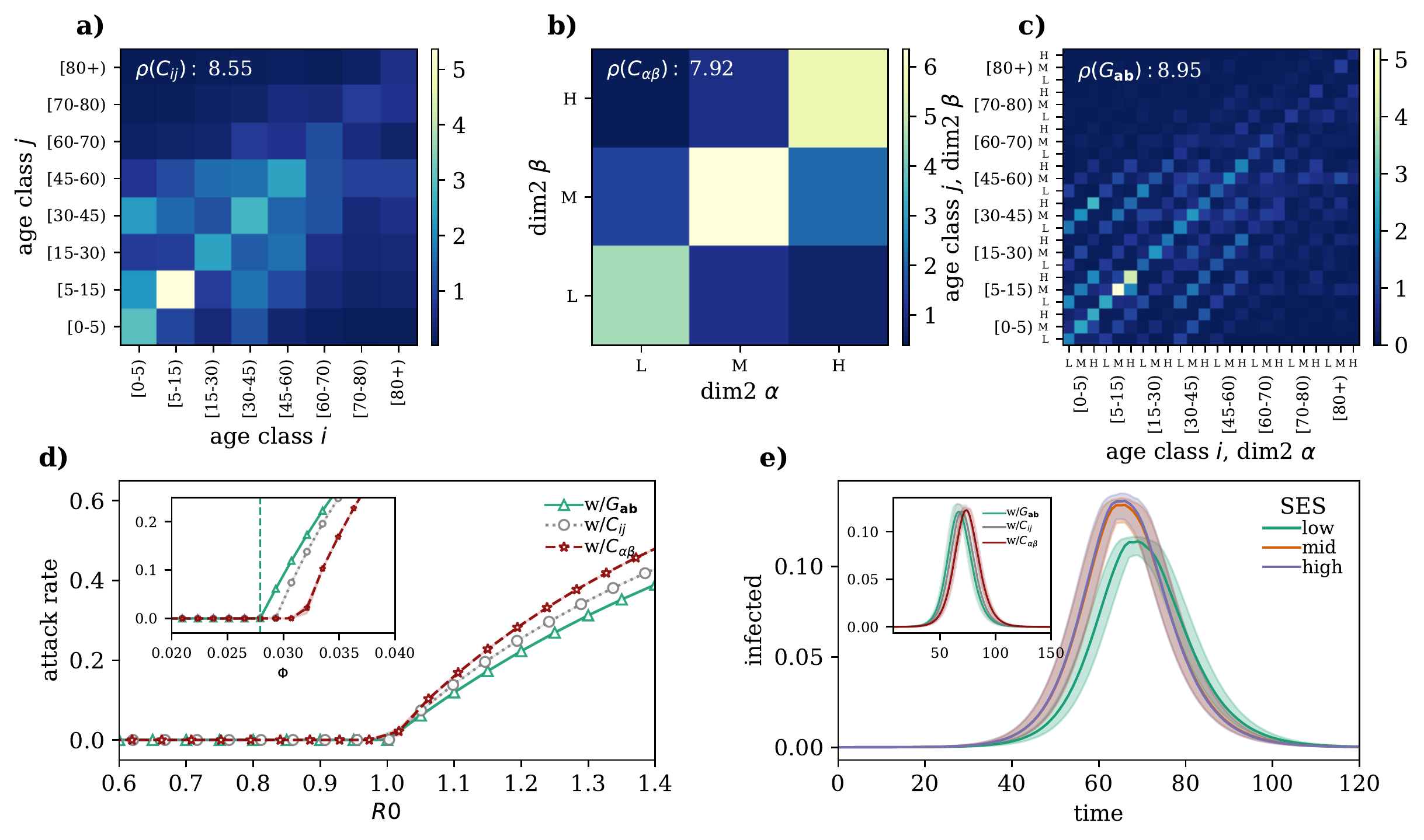}
    \caption{\textbf{Real world generalized contact matrices}. (a) Age contact matrix $\mathbf{C}$ ($8 \times 8$); (b) Socio-economic status (SES) contact matrix $C_{\alpha \beta}$ ($3 \times 3$) and (c) Generalized contact matrix with age and socio-economic status $\mathbf{G}$ ($24 \times 24$); (d) attack rate as a function of $R_0$ and (d) disease prevalence as a function of time. In panels (a-c) $\rho$ indicates the corresponding spectral radius. Results in panels (d) and (e) refer to the median of $500$ runs. Epidemiological parameters: $\Gamma=0.25, \Psi=0.4$, and $R_0=2.7$. Simulations start with $I_0= 100$ initial infectious seeds. The matrices have been computed using the contact diaries coming from the MASZK data collected in Hungary during June 2022.}
    \label{panel_4}
\end{figure*}

We applied the model to real data describing social contacts stratified by age and one SES variable (self-perceived wealth with respect to the average) in Hungary. The data has been collected via computer assisted surveys from $1,000$ respondents describing a representative sample of the Hungarian adult population in terms of gender, age, education level and type of settlement~\cite{koltai2022reconstructing}. We refer the reader to the Materials and Methods section and to the Supplementary Information for more details about the data and its collection. In Fig.~\ref{panel_4}-a we show the traditional contact matrix that focuses only on age. Though similar, the structure of contact patterns differs with respect to the equivalent matrix for the same country shown above in Fig.~\ref{panel_2}-a. This discrepancy is due to the different period these matrices represent. While the matrix in Fig.~\ref{panel_2}-a has been collected for the pre-COVID-19 period, data in Fig.~\ref{panel_4}-a records typical contact patterns during the COVID-19 pandemic, in particular in June 2022. Although at that time the vaccination campaign was in full swing and the number of confirmed cases and deaths was relatively low, the previous wave had peaked only a few months before and some level of social distancing was still in place (for more details about the construction and normalization of this matrix see the Supplementary Materials). The matrix confirms that children and young adults are the most active and that their interactions are largely assortative. However, the matrix features a rather smaller spectral radius with respect to the pre-pandemic contact matrix. This highlights the reduction in contacts caused by the COVID-19 emergency. In Fig.~\ref{panel_4}-b, we show the stratification of contacts considering only the SES variable, which divides the population into three SES classes with Low, Medium, and High wealth. Similar to other studies~\cite{leo2016socioeconomic,dong2020segregated}, we observe a strong diagonal component, confirming high levels of assortativity within SES groups. The population belonging to mid-low SES features lower assortativity and activity. Furthermore, the off-diagonal elements display higher levels of segregation for people in the low SES bracket, as individuals in mid and high SES tend to interact more with each other. In Fig.~\ref{panel_4}-c we show the flattened representation of the generalized contact matrix. Its spectral radius is higher than the other two, but the difference is clearly more limited with respect to the synthetic case we discussed in the previous section. In Fig.~\ref{panel_4}-d we show the attack rate as a function of $R_0$ for a disease spreading in a susceptible population described with each of the three contact matrices discussed. 
It is important to stress that these simulations do not aim to reproduce the evolution of the COVID-19 pandemic in Hungary. Instead, they describe a hypothetical disease spreading on different representations of the contact patterns measured in June 2022 to showcase the possible variations in the description of an epidemic. 
The theoretical critical value for the basic reproductive number (being at $R_0=1$) well captures the numerical simulations. Furthermore, for a given value of $R_0>1$ the attack rate estimated with a model featuring the generalized contact matrix (green triangles) is always smaller than any of the two other models. Also in this case, the more realistic description of contacts across multiple dimensions constrains the unfolding of the disease with respect to scenarios that neglect one of the two dimensions. However, the model featuring the contact matrix stratified by SES only (red stars) leads to higher attack rates than the model featuring an age-stratified contact matrix. 
The inset  confirms the validity of the mathematical formulation and highlights how a higher spectral radius results in a lower critical value for $\Phi$. In Fig.~\ref{panel_4}-e, we plot the incidence of the disease for the model with generalised contact matrix, splitting the three SES groups. Interestingly, we observe that individuals in the high SES group are the most affected in terms of contracted infections. This is due to their higher contact activity. Those in the mid-SES follow very closely, partially due to the strong interaction activities with the first group. Individuals in the low SES group, instead, are affected later and with a lower incidence rate. In this scenario, where the population is not subject to further NPIs nor spontaneously changing behavior during the epidemic, the higher level of segregation experienced by the low SES group has the silver lining effect of shielding that group from the epidemic, in accordance with empirical observations in the country~\cite{oroszi2022characteristics}. In the inset of Fig.~\ref{panel_4}-e we show the overall evolution of the disease obtained by modeling the epidemic with each one of the three contact matrices. The plot highlights how the chosen representation of contacts influences the description of the disease. Considering different numbers or types of dimensions, in this case, induces differences in the estimation of peak time, which is a key variable used to inform public health measures. In the Supplementary Information, we repeat the same analyses considering data collected at different time points of the COVID-19 pandemic, in April and November $2021$. While the overall results confirm the picture that emerged here, models fed with standard and generalized contact matrices lead to estimations of the attack rates, for a given $R_0$, which are closer to each other. Nevertheless, the model featuring generalized contact matrices allows capturing heterogeneities in the incidence across SES groups, which are invisible to standard approaches.

\section*{Discussion}

In this study, we developed a novel mathematical epidemic modelling framework featuring \emph{generalized contact matrices}, that is, contact matrices that are stratified according to multiple dimensions such as variables linked to socio-economic status (SES). The goal of our work is to move beyond the traditional representation of mixing patterns, based on age-specific or context-specific contact rates, allowing for the development of structured epidemic models that incorporate multiple groups of the population at the same time. 
First, we showed that the basic reproductive number, $R_0$, of a model featuring generalized contact matrices can still be obtained analytically via the next-generation matrix approach developed for traditional compartmental models~\cite{diekmann2010construction,blackwood2018introduction}. 
We then formally proved how the generalization to multiple dimensions leads to a value of $R_0$, which is necessarily equal to or larger than for models considering only one grouping dimension. Interestingly, we found how non-trivial correlations of contacts across and within groups, such as assortativity and heterogeneous activity, increase the value of $R_0$. Finally, through numerical simulations, we showed how models featuring generalized contact matrices better capture heterogeneous behaviours across population subgroups, such as varying rates of adherence to NPIs, as observed in reality~\cite{weill2020social, valdano2021highlighting, gauvin2021socio,gozzi2021estimating}.  
By using both synthetic and real data from Hungary, we demonstrated how neglecting additional dimensions in favour of simpler representations of social contacts may result in large misrepresentations of attack rates, epidemic thresholds, and disease burden across different groups of a population.  

The COVID-19 pandemic has tragically reminded us that health emergencies do not affect populations equally and that social, economic, and cultural forces fundamentally shape the outcome of disease outbreaks, reflecting the socio-economic inequalities of the society at large~\cite{karmakar2021association, zelner2021racial, siu2021health, sweeney2021exploring}. These observations are in stark contrast with the traditional structure of epidemic models, which generally neglect all variables but age as the key demographic feature defining disease transmission in a population~\cite{keeling08}. Recent studies have highlighted the urgent need for novel modelling approaches that can account for the multiple social dimensions that define the risk of infection and infection outcomes~\cite{zelner2022there, buckee2021thinking,tizzoni2022addressing}. In the wake of the COVID-19 pandemic, works have introduced social contact matrices stratified by alternative demographic groups, such as racial and ethnic subpopulations~\cite{ma2021modeling}, but usually without considering more than one dimension at once. As an alternative, scholars have developed large-scale individual-based models that reconstruct synthetic populations of millions of agents with extreme realism, including many key socio-economic characteristics~\cite{chang2021mobility, aleta2022quantifying, aleta2020modelling, pangallo2022unequal}. Such models, however, require the availability of highly resolved mobility data derived e.g., from mobile phone logs, and large computational infrastructures for data processing and simulations. Previous studies have compared different data representation methods, ranging from fully homogeneous mixing to temporal networks, to identify the relevant trade-offs between compactness and realism~\cite{machens2013infectious, aleta2020data, miller2013incorporating}. Our work attempts to strike a balance between traditional approaches that are too simplistic and the complexity of large-scale synthetic populations, providing a general and flexible scheme to define epidemic models with multiple interacting population subgroups. 

As for any structured epidemic modelling approach, our model requires data to parameterize social contact rates across subgroups. Social contact surveys have been and will continue to be an important asset for the study of mixing behaviors both in normal conditions and during epidemic outbreaks~\cite{mossong2008social, klepac2018contagion, melegaro2013_zimb, mousa2021social, ajelli2017estimating, zhang2020changes, gimma2022changes}. Future contact surveys could include additional demographic and socio-economic dimensions beyond the usual age/gender components. The work we presented here shows how such dimensions can be easily integrated by generalizing traditional epidemic models. In some cases, large-scale contact surveys including several population attributes can be unfeasible, especially in resource-poor settings. Alternative approaches could infer contact patterns from the analysis of demographic data that are routinely collected by census surveys~\cite{Mistry2021, fumanelli2012inferring}. Other proxies of contact rates, derived from mobility data, could be similarly used to infer contact patterns among different socio-economic groups, as demonstrated by recent studies focusing on experienced segregation in large US metropolitan areas~\cite{moro2021mobility, yabe2023behavioral}.  

It is important to mention the limitations of our work and the directions for future developments of our study. Admittedly, the model we developed to generate synthetic generalized contact matrices does not attempt to reproduce empirical observations from real data, but rather to provide a general test bed for investigation. This choice was guided by the lack of available data about the stratification of contacts across multiple dimensions in different countries. As described, we had access to real-world data only for Hungary. Arguably, one country is not enough to develop a general model and more work is needed to design and test other approaches. Similarly, the methodology we used to simulate the effects of NPIs was driven by simplicity rather than realism. Our intention was to showcase the potential of our model to capture possible heterogeneities in behaviours rather than reality. Data from Hungary contained limited information about social contacts of children. Hence, we had to introduce a few assumptions about their structure. Another important assumption we made is about the independence of age and other socio-economic dimensions. While this assumption was necessary to showcase our methodology, future work is needed to investigate the impact of such correlations on the description of epidemic processes and to explore the applicability/extension of methods for dimensionality reduction. 

Finally, it is important to acknowledge important ethical and privacy concerns linked with data collection efforts that inspect individuals about several variables. Indeed, by increasing the number of dimensions the size of each surveyed group rapidly decreases. Risks of re-identification are real and of particular concern, especially for minorities and underrepresented groups~\cite{rocher2019estimating}. Progresses in the direction we have proposed here are necessarily linked to progress in data collection and dissemination that reconcile the development of better public health tools on the one side with privacy rights on the other~\cite{oliver2020mobile}. Arguably, it is this unresolved tension that, among other reasons, has so far limited the collection and sharing of contact surveys data including more dimensions besides age. Nevertheless, the data we use demonstrates that such data collection is possible in an anonymous and privacy protected way.

Overall, our work contributes to the literature by attempting to bring socio-economic and other dimensions to the forefront of epidemic modelling. Tackling this issue is crucial for developing more precise descriptions of epidemics, and thus to design better strategies to contain them.

\section*{Materials and Methods}

\subsection*{Synthetic generalized contact matrices} 

We developed a model to build synthetic generalized contact matrices $G_{\mathbf{a},\mathbf{b}}$. For simplicity and clarity of exposition let us consider here only two dimensions: age and an additional variable as for example one SES indicator. In these settings $\mathbf{a}=(i,\alpha)$ and $\mathbf{b}=(j,\beta)$. The indices $i$ and $j$ refer to the age group while $\alpha$ and $\beta$ to the SES of the ego and the alter respectively. The model takes as input a real contact matrix $C_{ij}$ describing the contact rates between age-brackets. As described above $R_{ij}=C_{ij}N_i$ is the total number of contact between the two groups in a given period. To build the generalized contact matrices $G_{\mathbf{a},\mathbf{b}}$ we first split the total contacts $R_{ij}$ across the second dimension and then we compute the contacts rates. In other words $R_{ij}=\sum_{\alpha, \beta}R_{i\alpha, j\beta}$. Furthermore, the model must respect a key symmetry of these matrices: $R_{i\alpha, j\beta}=R_{j\beta, i\alpha}$. In words, the number of contacts that individuals in age group $i$ and SES $\alpha$ have with individuals in age group $j$ and SES $\beta$, must be equal to the number of contacts that individuals in age group $j$ and SES $\beta$ have with individuals in age group $i$ and SES $\alpha$. Hence $R_{i\alpha,j\beta}=R_{i\alpha,j\beta}^\top$. This property implies that for all $i=j$ $R_{i\alpha,i\beta}=R_{i\beta,i\alpha}$. In general this is not the case for $i\neq j$. For a given pair $i$ and $j$ the matrix $R_{i\alpha,j\beta}$ is of size $V_1\times V_1$. Hence, for any $i\neq j$, we need a model to set $W=V_1^2-1$ elements of the matrix. For all $i=j$ instead, the symmetry of the matrix is such that we need to set only $Y=V_1+ \frac{V_1(V_1-1)}{2}-1$ elements. We define the $W$ and $Y$ values of the matrices via simple model where, for any $i$ and $j$ pair, i) each SES $\alpha$ is responsible for a fraction $P_\alpha$ of $R_{ij}$ connections, ii) a fraction $q_\alpha$ of these are on the diagonal (i.e., in-group connections) and $1-q_\alpha$ are instead off-diagonal. This parameter controls the assortativity of connections within each group. The fractions $P_\alpha$ and $q_\alpha$ are input parameters. In case $W$ or $Y$ are equal to this number ($2V_1-1$), the constraints imposed by our assumptions allow to define all the entries of the matrix. If instead $W>2V_1-1$, other $W-2V_1+1$ free parameters are required for each $i$ and $j$ pair. Interestingly, in case $P_\alpha=q_\alpha=\frac{N_\alpha}{N}$ it is easy to show how the contacts between groups become the expectation value of a random mixing process. We refer the reader to the Supplementary Information for more details about synthetic generalized contact matrices.

\subsection*{Real generalized contact matrices}
We built generalized contact matrices stratified in two dimensions by using real data from the \textit{MASZK} study~\cite{karsai2020hungary,koltai2022reconstructing} (for more information about the data set see Supplementary Materials). The data provided us with a range of information about the anonymous participants including their perceived wealth with respect to the average, which is one SES variable we rely on. Information on social interactions have been collected in two different ways i) in an aggregate form, such that each participant reported the number of contacts they had with individuals in each of the eight age bracket considered, ii) in a diary in which each participant listed one by one the interactions they had on a given day by reporting some meta information of the \textit{contactee} such as their age and SES. In particular, the average number of contacts of an individual in age class $i$, and SES $\alpha$ with an individual in age class $j$, and SES $\beta$ is $G_{\mathbf{a,b}}$ where $\mathbf{a}=[i,\alpha]$ and $\mathbf{b}=[j,\beta]$. However, the \textit{MASZK} data provided us with diary information only for the adult population (individuals older than 15 years old). For the children, their average number of contacts is available only in the aggregate form for the eight age brackets considered. Thus, we assigned the average number of contactee to the different SES as follow: $G_{i\alpha, j \beta} = G_{i \alpha, j} u_{\alpha \beta}$, where $G_{i \alpha, j}$ is the average number of contacts that individual of age group $i$ and SES $\alpha$ has with all the individuals of age group $j$, and $u_{\alpha \beta}$ is a parameter that controls how these contacts are distributed among individuals of different SES.

\subsection*{Numerical simulations} To investigate the effect of the generalized contact matrices $G_{\mathbf{a},\mathbf{b}}$ on infection transmission dynamics, we developed a stochastic, discrete-time, compartmental model where the transitions among compartments are simulated through chain binomial processes. In particular, at time step $t$ the number of individuals in group $\mathbf{a}$ and compartment $X$ transiting to compartment $Y$ is sampled from $PrBin(X\mathbf{a}(t),p_{X\mathbf{a}\xrightarrow{}Y\mathbf{a}}(t))$,where $p_{X\mathbf{a}\xrightarrow{}Y\mathbf{a}}(t)$ is the transition probability.

\paragraph{Acknowledgements}
The authors gratefully thank Ciro Cattuto and Alessandro Vespignani for useful discussions. A.M. and M.K. were supported by the Accelnet-Multinet NSF grant. A.M. is grateful for the support from CEU and NetSI (Northeastern University). M.K. acknowledges support from the ANR project DATAREDUX (ANR-19-CE46-0008); the SoBigData++ H2020-871042; the EMOMAP CIVICA projects; and the National Laboratory for Health Security, Alfréd Rényi Institute, RRF-2.3.1-21-2022-00006. L.D. acknowledges support from the Fondation Botnar and from the Lagrange Project of the ISI Foundation funded by CRT
Foundation.

\paragraph{Authors Contribution} 
A.M. performed the numerical simulations and data analysis. A.M., L.D., and N.P. developed the analytical formulation. A.M., M.T., and N.P. wrote the first draft of the manuscript. All authors designed the study, interpreted the results, edited and approved the manuscript. To whom correspondence should be addressed. E-mail: n.perra@qmul.ac.uk

\bibliography{refs}

\clearpage
\section*{Supporting Information}

\subsection*{Epidemic models with age-stratified contact matrices}

Standard approaches to model the spreading of infectious diseases often acknowledge the stratification of contacts across age-brackets. To this end, contact matrices $\mathbf{C}$ are introduced. The element $C_{ij}$ quantifies the average number of contacts that an individual in age-bracket $i$ has with individuals in age group $j$ within a certain time window~\cite{mossong2008social,prem2017projecting,Mistry2021}. The population hence, it is divided into age brackets so that $N=\sum_{i=1}^{K}N_i$. 
The variables $N_i$ capture the number of individual in age group $i$ while $K$ indicates the number of different age-groups. Given the definition of contact matrices, we can define $R_{ij}=C_{ij}N_i$ as the total number of contacts that individuals in $i$ have with those in $j$. This matrix is clearly symmetric: $R_{ji}=R_{ij}$. However since in general $N_i\neq N_j$ the entries of the contact matrices are not, i.e., $C_{ij}=R_{ij}N_i\neq C_{ji}=R_{ji}N_j$.

Let us now consider a disease whose natural history can be described with a Susceptible-Exposed-Infected-Recovered model~\cite{keeling08}. The epidemic dynamics are encoded in the following set of differential equations:
\begin{eqnarray}
d_t S_{i}(t) &=& -\Lambda_i (t) S_i(t), \nonumber \\
% d_t E_i(t) &=& \Lambda S_i(t) -\Psi E_i(t), \nonumber \\
d_t E_i(t) &=& \Lambda_i(t) S_i(t) -\Psi E_i(t), \nonumber \\
d_t I_i(t) &=& \Psi E_i(t)-\Gamma I_i (t),\nonumber \\
d_t R_i(t) &=& \Gamma I_i(t).
\end{eqnarray}
Susceptible individuals, in contact with infected, might be exposed to the virus with a rate driven by the force of infection $\Lambda_i(t)$; exposed are not yet infectious and transition to the infected compartment with rate $\Psi$; infected individuals recover with rate $\Gamma$. The force of infection is then defined as the per-capita rate at which susceptibles are exposed to the disease:
\begin{equation}
\label{force}
    \Lambda_i(t)= \Phi \sum_{j=1}^{K}C_{ij}\frac{I_{j}(t)}{N_j}
\end{equation}
where $\Phi$ is the transmissibility of the disease and the temporal dependence is induced by the variation in the number of infected across age brackets.

\subsection*{Derivation of $R_0$}
A fundamental quantity to understand the epidemic dynamics is the basic reproductive number, $R_0$, defined as the number of secondary infections generated by a single infected individual in a otherwise susceptible population~\cite{keeling08}. 
The basic reproductive number is function of the features of the disease and the contact patterns of the population. 
The next generation matrix approach can be used to determine a closed form expression of $R_0$. While we refer the reader to Ref.~\cite{blackwood2018introduction} for an overview of the method, in the following we provide a summary of its derivation. To keep equations more readable, we henceforth drop the explicit time dependence notation.

The first step is to focus our attention only on the compartments that describe any stages of the infection: $E_i$ and $I_i$ $i \in [1,K]$ in our case. It is convenient to re-write the differential equations for these compartment as 
\begin{equation}
    d_t \mathbf{x}=f(\mathbf{x})-w(\mathbf{x})
\end{equation}
where $\mathbf{x} \in [E_1, \ldots E_K, I_1, \ldots I_K]$, $f(\mathbf{x})$ encodes all terms that lead to new infections and $w(\mathbf{x})$ all other transitions in and out of the compartments. More explicitly, we can write
\begin{equation}
   d_t \mathbf{x}= \begin{bmatrix}
d_t E_1 \\
\vdots \\
d_t E_K \\
d_t I_1 \\
\vdots \\
d_t I_K \\
\end{bmatrix}= f(\mathbf{x})-w(\mathbf{x})=\begin{bmatrix}
\Lambda_1 S_1 \\
\vdots \\
\Lambda_K S_K\ \\
0 \\
\vdots \\
0 \\
\end{bmatrix}-\begin{bmatrix}
\Psi E_1 \\
\vdots \\
\Psi E_K \\
\Gamma I_1-\Psi E_1 \\
\vdots \\
\Gamma I_K-\Psi E_K\\
\end{bmatrix}
\end{equation}
As alluded above, the expression of $R_0$ is linked to the early epidemic dynamics which can be linearized by calculating the Jacobian at the disease free equilibrium (DFE) $(S^*_{i},E^*_{i},I^*_{i},R^*_{i})=(N_{i},0,0,0)$ for all age groups. Using the Jacobian matrix $\mathbf{J}$ (which has size $2K\times 2K$) at the DFE we can write 
\begin{equation}
 d_t \mathbf{x}=\mathbf{J}\mathbf{x}   
\end{equation}
This expression can be conveniently factorized as $d_t \mathbf{x}=(\mathbf{F}-\mathbf{W})\mathbf{x}$ where the matrix $\mathbf{F}$ is Jacobian applied to $f(\mathbf{x})$ and similarly the matrix $\mathbf{W}$ is the Jacobian applied to $w(\mathbf{x})$. In particular, we have:
\begin{equation}
\footnotesize
    \mathbf{F}=\begin{bmatrix}
\partial_{E_1}(\Lambda_1 N_1) & \ldots  & \partial_{E_K}(\Lambda_1 N_1) &  \partial_{I_1}(\Lambda_1 N_1) & \ldots & \partial_{I_K}(\Lambda_1 N_1)\\
\partial_{E_1}(\Lambda_2 N_2) & \ldots  & \partial_{E_K}(\Lambda_2 N_2) &  \partial_{I_1}(\Lambda_2 N_2) & \ldots & \partial_{I_K}(\Lambda_2 N_2) \\
\vdots & \vdots & \vdots & \vdots & \vdots & \vdots\\
\partial_{E_1}(\Lambda_K N_K) & \ldots  & \partial_{E_K}(\Lambda_K N_K) &  \partial_{I_1}(\Lambda_K N_K) & \ldots & \partial_{I_K}(\Lambda_K N_K)\\
0  & \ldots  & 0 & 0 & \ldots  & 0 & \\
0  & \ldots  & 0 & 0 & \ldots  & 0 &\\
\vdots  & \vdots & \vdots & \vdots & \vdots & \vdots\\
0 & \ldots  & 0 & 0 & \ldots  & 0 & \\
\end{bmatrix}=\begin{bmatrix}
0 & \ldots  & 0 &  \partial_{I_1}(\Lambda_1 N_1) & \ldots & \partial_{I_K}(\Lambda_1 N_1)\\
0 & \ldots  & 0 &  \partial_{I_1}(\Lambda_2 N_2) & \ldots & \partial_{I_K}(\Lambda_2 N_2) \\
\vdots & \vdots & \vdots & \vdots & \vdots & \vdots\\
0 & \ldots  & 0 &  \partial_{I_1}(\Lambda_K N_K) & \ldots & \partial_{I_K}(\Lambda_K N_K)\\
0  & \ldots  & 0 & 0 & \ldots  & 0 & \\
0  & \ldots  & 0 & 0 & \ldots  & 0 &\\
\vdots  & \vdots & \vdots & \vdots & \vdots & \vdots\\
0 & \ldots  & 0 & 0 & \ldots  & 0 & \\
\end{bmatrix}
\end{equation}
We stress how the components of each gradient, are computed at the DFE, i.e., $S_i \rightarrow N_i$. By looking at Eq.~\ref{force} we note how each partial derivative in $p$ selects the $p$-th element of the sum hence:
\begin{equation}
    \mathbf{F}=\Phi \begin{bmatrix}
0 & \ldots  & 0 &  \frac{C_{11}}{N_1}N_1 & \ldots & \frac{C_{1K}}{N_K}N_1\\
0 & \ldots  & 0 &  \frac{C_{21}}{N_1}N_2 & \ldots & \frac{C_{2K}}{N_K}N_2\\
\vdots & \vdots & \vdots & \vdots & \vdots & \vdots\\
0 & \ldots  & 0 &  \frac{C_{K1}}{N_1}N_K & \ldots & \frac{C_{KK}}{N_K}N_K\\
0 &\ldots  & 0 & 0 & \ldots  & 0 & \\
0 & \ldots  & 0 & 0 & \ldots  & 0 &\\
\vdots & \vdots & \vdots & \vdots & \vdots & \vdots\\
0 & \ldots  & 0 & 0 & \ldots  & 0 & \\
\end{bmatrix}=\Phi \begin{bmatrix}
\mathbf{0} & \mathbf{\tilde{C}} \\
\mathbf{0} &\mathbf{0} 
\end{bmatrix}
\end{equation}
where we denoted with $\mathbf{0}$ $K\times K$ blocks of zeros and the generic entry of $\mathbf{\tilde{C}}$ is $\tilde{C_{ij}}=\frac{C_{ij}}{N_j}N_i$. Using the same method we can easily write the expression of the matrix $\mathbf{W}$ as:
\begin{equation}
    \mathbf{W}=\begin{bmatrix}
\Psi & \ldots  & 0 &  0 & \ldots  & 0 \\
\vdots & \vdots & \vdots & \vdots & \vdots & \vdots\\
0 & \ldots  & \Psi & 0 & \ldots  & 0 \\
-\Psi & \ldots  & 0 & \Gamma & \ldots  & 0 & \\
\vdots & \vdots & \vdots & \vdots & \vdots & \vdots\\
0 & \ldots  & -\Psi & 0 & \ldots  & \Gamma & \\
\end{bmatrix}= \begin{bmatrix}
\Psi\mathbf{I} & \mathbf{0} \\
-\Psi\mathbf{I} & \Gamma\mathbf{I} 
\end{bmatrix}
\end{equation}
where $\mathbf{I}$ are identity matrices of size $K\times K$. The expression for $R_0$ is linked to the two matrices as $R_0=\rho(\mathbf{FW}^{-1})$, where $\rho$ denotes the spectral radius. It is easy to show how
\begin{equation}
    \mathbf{FW}^{-1}=\Phi \begin{bmatrix}
\mathbf{0} & \mathbf{\tilde{C}} \\
\mathbf{0} &\mathbf{0} 
\end{bmatrix}\begin{bmatrix}
\frac{1}{\Psi}\mathbf{I} & \mathbf{0} \\
\frac{1}{\Gamma}\mathbf{I} & \frac{1}{\Gamma}\mathbf{I} 
\end{bmatrix}=\frac{\Phi}{\Gamma} \begin{bmatrix}
\mathbf{\tilde{C}} & \mathbf{\tilde{C}} \\
\mathbf{0} &\mathbf{0} 
\end{bmatrix}
\end{equation}
hence, we finally can write
\begin{equation}
    R_0=\frac{\Phi}{\Gamma} \rho(\mathbf{\tilde{C}})
\end{equation}
Contact matrices are often stratified also for the context (i.e., location $l$) where interactions take place~\cite{Mistry2021}. The entries $C_{ij}$ are then expressed as
\begin{equation}
    C_{ij}=\sum_{l}\omega_l C_{ij}^{(l)}
\end{equation}
where $\omega_l$ are weights capturing possible heterogeneities in the relevance of contacts in each context in the transmission of the disease. In this formulation, the inclusion of the different contexts does not change the expression of $R_0$ as the $\omega_l$ are assumed to be homogeneous across age-brackets.

\subsection*{Generalized contact matrices}
\label{Gab}

In this work, we shift from the standard contact matrices that stratify contacts for age and possibly context, to a generalized version that allows for $m$ other dimensions. Here, we focus the attention towards socio-economic status (SES) variables as additional dimensions. However, the framework is general and can be applied to any categorical variable of epidemiological interest. \\
In more details, we describe the generalized contact matrices as $G_{\mathbf{a},\mathbf{b}}$, where $\mathbf{a}=(i,\alpha, \beta,\ldots,\gamma)$ and $\mathbf{b}=(j,\eta, \mu,\ldots,\xi)$ are tuples (i.e., index vectors) representing individuals membership to each category. We note how we adopted Greek letters for the additional dimensions, though not necessary. With these matrices we can, for example, capture contacts stratification according to age, income ($\alpha$), and education attainment   ($\beta$). In this scenario $G_{\mathbf{a}, \mathbf{b}}$ would describe the average number of contacts that an individual in age bracket $i$, income $\alpha$, and education $\beta$ has with people in age group $j$, income $\eta$, and education $\mu$ in a given time window. From this perspective, the matrix $\mathbf{C}$ can be thought as an aggregation of contacts patterns at a lower levels of stratification. In other words, the standard matrices can be viewed as aggregation along all dimensions affecting the organization of contacts but age. \\
The total number of contacts of individuals in a given age group $i$ with other in $j$ can be written as $R_{ij}=C_{ij}N_i$. Considering the generalized contact matrices we have $C_{ij}N_i=\sum_{\mathbf{a' b'}}G_{\mathbf{a'},\mathbf{b'}}N_{\mathbf{a'}}$, where $\mathbf{a'}=\mathbf{a}-\{i\}$ and $\mathbf{b'}=\mathbf{b}-\{j\}$ are the index vectors $m$ capturing all dimensions but age. Since the matrices $\mathbf{G}$ has $m+1$ dimensions, we can aggregate contacts in $m+1$ possible ways, for example computing the average contacts that individual in one of the SES $\alpha$ has with others in same SES $\beta$ obtaining
\begin{equation}
C_{\alpha \beta}N_\alpha=\sum_{\mathbf{a'b'}}G_{\mathbf{a'},\mathbf{b'}}N_{\mathbf{a'}}
\end{equation}
where now $\mathbf{a'}=\mathbf{a}-\{\alpha\}$ and $\mathbf{b'}=\mathbf{b}-\{\beta\}$. In general, $C_{ij}\neq C_{\alpha \beta}$. Indeed, the number of groups and the number of individuals in each of them might be different. Looking at contacts patterns from multiple dimensions of stratification allows us to observe how the way interactions are aggregated (i.e., by age or other dimensions) might affect the estimation of key epidemiological parameters and of the dynamics.\\

We denote with $K$ the number of age groups, while with $V_p$ the number of groups in each of the $m$ other dimensions (i.e., $p \in [1,m]$). While the generalized matrix $\mathbf{G}$ can be naturally described as a multidimensional matrix, the use of $T=K\prod_{p=1}^{m}V_p \times K\prod_{p=1}^{m}V_p$ index vectors allows for flattened representation in a squared bi-dimensional matrix of size $K\prod_{p=1}^{m}V_p \times K\prod_{p=1}^{m}V_p$. We note how the formulation can easily consider different contexts where interactions take places, i.e., $G_{\mathbf{a},\mathbf{b}}=\sum_{l}\omega_l G^{(l)}_{\mathbf{a},\mathbf{b}}$ where $\omega_l$ captures the relative importance of the different social settings in the transmission~\cite{Mistry2021}.\\

\subsection*{Synthetic generalized contact matrices}
\label{sec_sy_G}

In this section, we provide details about the generation of synthetic generalized contact matrices we built to explore the possible effects of different mixing patterns among individuals. For simplicity we first consider only two dimensions: age and a socio-economic status (SES) variable (though any other categorical variable could be used).

We developed a model to derive $G_{\mathbf{a},\mathbf{b}}$ where $\mathbf{a}=(i,\alpha)$ and $\mathbf{b}=(j,\beta)$, and $i$ and $j$ refers to the age group while $\alpha$ and $\beta$ to the SES of the ego and the alter respectively. We start from a real contact matrix $C_{ij}$ describing the contact rates between age-brackets $i$ and $j$. We can define $R_{ij}=C_{ij}N_i$ as the total number of contacts between the two groups in the a given period. While the matrix $C_{ij}$ is not symmetric, $R_{ij}$ is since the number of \emph{raw} contacts between two groups is the same, i.e., $R_{ij}=C_{ij}N_i=R_{ji}=C_{ji}N_j$. To build the generalized contact matrices $G_{\mathbf{a},\mathbf{b}}$ we first split the total contacts $R_{ij}$ across the second dimension and then we compute the contacts rates. In other words $R_{ij}=\sum_{\alpha, \beta}R_{i\alpha, j\beta}$. The problem is then how to get values for the elements $R_{i\alpha, j\beta}$. It is useful to reflect on the properties of these matrix. First, $R_{i\alpha, j\beta}=R_{j\beta, i\alpha}$. In words, the number of contacts that individuals in age group $i$ and SES $\alpha$ have with individuals in age group $j$ and SES $\beta$, must equal the number of contacts that individuals in age group $j$ and 
SES $\beta$ have with individuals in age group $i$ and SES $\alpha$. This property implies that $R_{i\alpha,j\beta}=R_{i\alpha,j\beta}^\top$. Indeed, for a given pair of indices $i$ and $j$, one can think about $R_{i\alpha,j\beta}$ as a $V_1\times V_1$ matrix (since $\alpha, \beta \in [1,\ldots, V_1]$), which describes the contact patterns between those two age groups across SES. The matrix obtained by switching the indices for age (i.e., $R_{j\alpha,i\beta}$) is the transpose of the first. This property has another implication: for all $i=j$ the matrix $R_{i\alpha,i\beta}$ must be symmetric: $R_{i\alpha,i\beta}=R_{i\beta,i\alpha}$. In general this is not the case for $i\neq j$. Second, as mentioned above the sum over all $\alpha$ and $\beta$ is set by the number of contacts between age group $i$ and $j$: $R_{ij}=\sum_{\alpha, \beta}R_{i\alpha, j\beta}$. 

Given these features, how can we set the entries of these matrices? As mentioned, for a given pair $i$ and $j$ the matrix $R_{i\alpha,j\beta}$ is of size $V_1\times V_1$ since $\alpha, \beta \in [1,\ldots, V_1]$. For any $i\neq j$, we need a model to set $W=V_1^2-1$ elements of the matrix (the minus one is due to the constraint introduced by the second property defined above). For all $i=j$ instead, the symmetry of the matrix is such that we need to set only $Y=V_1+ \frac{V_1(V_1-1)}{2}-1$ elements. The first $V_1$ come from the diagonal (i.e., $\alpha=\beta$), the $\frac{V_1(V_1-1)}{2}$ are instead the off-diagonal values, the minus one is due to the constraint of the total number of contacts which is set by $R_{ij}$. To set the $W$ and $Y$ values of the matrices we consider that for any $i$ and $j$ pair: 
\begin{enumerate}
    \item each SES $\alpha$ is generally responsible for $P_\alpha$ fraction of $R_{ij}$ connections, where $\sum_{\alpha}P_{\alpha}=1$
    \item we assume that a fraction $q_\alpha$ of these are on the diagonal (i.e., in-group connections) and $1-q_\alpha$ are instead off-diagonal. This parameter controls the assortativity of connections within each group.
\end{enumerate}

In what follows, we consider the fractions $P_\alpha$ and $q_\alpha$ as input parameters. In case $W$ or $Y$ are equal to this number ($2V_1-1$), the constraints imposed by our assumptions allow to define all the entries of the matrix. If instead $W>2V_1-1$, other $W-2V_1+1$ parameters are required.

Let us consider first the case in which  $V_1=3$, $\alpha, \beta \in [1,2,3]$. In these settings $W=8$, and $Y=5$ which is equal to $2V_1-1$. Hence, defining the two set of input parameters is enough to obtain the matrix for any $i=j$, while still $3$ additional parameters are needed for $i\neq j$. It is important to stress how the first property described above implies that we need to define only this number for all $i<j$. The correspondent values for $i>j$ are readily obtained just transposing the matrices. It is useful, to split the two cases:\\

\textbf{Case $i=j$.}\\

For all $i=j$ the matrices $R_{i\alpha,i\beta}$ take the form of

\begin{equation}
   R_{i\alpha,i\beta}=R_{ii}\left(
\begin{matrix}
q_1 P_1 & p_{i1,i2}  & p_{i1,i3} \\
p_{i1,i2} & q_2 P_2 & p_{i2,i3} \\
p_{i1,i3}& p_{i2,i3}  & q_3 P_3 \\
\end{matrix}
\right)
\end{equation}
where each $p_{i\alpha,i\beta}\in [0,1]$ is defined such that $R_{i\alpha,i\beta}=R_{ii}p_{i\alpha,i\beta}$. As the matrix needs to be symmetric we have written, for example, $p_{i1,i2}$ instead of $p_{i2,i1}$. Due to our assumptions the sum of the off-diagonal elements of each rows should sum to $(1-q_\alpha)P_\alpha=\Phi_\alpha$, while on the diagonal we have $q_\alpha P_\alpha$. Thus the following system of equations must be respected:
\begin{equation}
\begin{cases}
    p_{i1,i2}+p_{i1,i3}=\Phi_1 \\
p_{i1,i2}+p_{i2,i3}=\Phi_2 \\
p_{i1,i3}+p_{i2,i3}=\Phi_3 \\
\end{cases}
\end{equation}
The solution, if any, is unique and can be written as 
\begin{equation}
\label{phat_v3}
    \mathbf{\hat{p}=A^{-1}\Phi}=\frac{1}{2}\left(
\begin{matrix}
\Phi_{1}+\Phi_{2}-\Phi_{3}\\
\Phi_{1}+\Phi_{3}-\Phi_{2}\\
\Phi_{2}+\Phi_{3}-\Phi_1
\end{matrix}
\right)
\end{equation}
where 
\begin{equation}
    \mathbf{A}=\left(
\begin{matrix}
1 & 1 &0\\
1 & 0 & 1\\
0 & 1 & 1
\end{matrix}
\right), \;\ \mathbf{\Phi}=\left(
\begin{matrix}
\Phi_{1}\\
\Phi_{2}\\
\Phi_{3}
\end{matrix}
\right), \;\ \mathbf{\hat{p}}=\left(
\begin{matrix}
p_{i1,i2}\\
p_{i1,i3}\\
p_{i2,i3}
\end{matrix}
\right)
\end{equation}

Let us consider now the case in which $V_1=4$. In this case, $Y=9$. As before, $2V_1-1$ parameters are inputs (three $P_\alpha$ and four $q_\alpha$). In this case, $Y>2V_1-1$. Hence, the system of $V_1$ differential equations in $V_1(V_1-1)/2=6$ (off-diagonal) elements required $Y-2V_1+1=2$ free parameters. Indeed, the matrix $R_{i\alpha,i\beta}$ can be written as: 

\begin{equation}
    R_{i\alpha,i\beta}=R_{ii}\left(
\begin{matrix}
q_1 P_1 & p_{i1,i2}  & p_{i1,i3} & p_{i1,i4} \\
p_{i1,i2} & q_2 P_2 & p_{i2,i3} & p_{i2,i4} \\
p_{i1,i3}& p_{i2,i3}  & q_3 P_3 & p_{i3,i4} \\
p_{i1,i4} & p_{i2,i4} & p_{i3,i4} & q_4 P_4
\end{matrix}
\right)
\end{equation}
Considering that the sum of each off-diagonal element in each row $\alpha $should sum to $(1-q_\alpha)P_\alpha$ we obtain a system of four equations: 

\begin{equation}
\begin{cases}
    p_{i1,i2}+p_{i1,i3}+p_{i1,i4}=\Phi_1 \\
p_{i1,i2}+p_{i2,i3}+p_{i2,i4}=\Phi_2 \\
p_{i1,i3}+p_{i2,i3}+p_{i3,i4}=\Phi_3 \\
p_{i1,i4}+p_{i2,i4}+p_{i3,i4}=\Phi_4 \\
\end{cases}
\end{equation}
As before, we can write the system of $6$ equations in matrix form as $\mathbf{A\hat{p}=\Phi}$ where:
\begin{equation}
    \mathbf{A}=\left(
\begin{matrix}
1 &1& 1 &0 &0& 0\\
1 &0& 0 &1 &1 &0\\
0& 1 &0 &1 &0 &1\\
0 &0 &1 &0 &1& 1
\end{matrix}
\right), \mathbf{\hat{p}}=\left(
\begin{matrix}
p_{i1,i2}\\
p_{i1,i3} \\
p_{i1,i4}\\
p_{i2,i3} \\
p_{i2,i4}\\
p_{i3,i4}
\end{matrix}
\right), \mathbf{\Phi}=\left(
\begin{matrix}
\Phi_{\alpha}\\
\Phi_{\beta}\\
\Phi_{\gamma}\\
\Phi_{\Gamma}
\end{matrix}
\right)
\end{equation}
Clearly the system is under-determined as it has four equations and six variables. To find the solution we need to fix two of such variables as for example $p_{i2,i4}$ and $p_{i3,i4}$. Doing so, we can write the system of equations as $\mathbf{A_r\hat{p}_r=\Phi_r}$ where
\begin{equation}
    \mathbf{A}_r=\left(
\begin{matrix}
1 &1& 1 &0\\
1 &0& 0 &1\\
0& 1 &0 &1\\
0 &0 &1 &0
\end{matrix}
\right), \mathbf{\hat{p}_r}=\left(
\begin{matrix}
p_{i1,i2}\\
p_{i1,i3} \\
p_{i1,i4}\\
p_{i2,i3} \\
\end{matrix}
\right), \mathbf{\Phi_r}=\left(
\begin{matrix}
\Phi_{1}\\
\Phi_{2}-p_{i2,i3}\\
\Phi_{3}-p_{i3,i4}\\
\Phi_{4}-p_{i2,i3}-p_{i3,i4}
\end{matrix}
\right)
\end{equation}
The subscript $r$ stands for reduced, as by fixing the $Y-2V_1+1$ variables we can move from a $4\times 6$ matrix to a $4\times 4$. The solution, if any, is now unique and can be written as:
\begin{equation}
\label{phat}
    \mathbf{\hat{p}_r=A^{-1}_r\Phi_r}
\end{equation}

For any general $V_1$, the method described above allows to define the matrix $R_{i\alpha, i\beta}$ passing through an under-defined system of linear equations. By setting the potential $Y-2V_1+1$ free parameters the solution, if it exists, is unique. However, we expect that some solutions might not physical. Indeed, the entries of the matrix must be all equal or larger than zero. Any solution that leads to negative values is not physical and must be neglected. This condition adds constraints to the possible values of the $Y$ parameters. It is important to notice how the free-parameters are in general function of $i$. For simplicity, and to avoid having to set $K\times(Y-2V_1+1)$ values across all age groups, could be set equal for all age groups.\\

\textbf{Case $i<j$.}\\

For all $i<j$ the matrices $R_{i\alpha,j\beta}$ take the form of

\begin{equation}
   R_{i\alpha,j\beta}=R_{ij}\left(
\begin{matrix}
q_1 P_1 & p_{i1,j2}  & p_{i1,j3} \\
p_{i2,j1} & q_2 P_2 & p_{i2,j3} \\
p_{i3,j1}& p_{i3,j2}  & q_3 P_3 \\
\end{matrix}
\right)
\end{equation}
where each $p_{i\alpha,j\beta}\in [0,1]$ is defined such that $R_{i\alpha,j\beta}=R_{ij}p_{i\alpha,j\beta}$. The difference with respect to the previous case is that now the matrix is not, in general, symmetric. As such, we need to find a model to set the $W=V_1^2-1$ values. However, as in the previous case, due to our assumptions, the sum of the off-diagonal elements of each rows should sum to $(1-q_\alpha)P_\alpha=\Phi_\alpha$, while on the diagonal we have $q_\alpha P_\alpha$. Thus the following system of equations must be respected:
\begin{equation}
\begin{cases}
    p_{i1,j2}+p_{i1,j3}=\Phi_1 \\
p_{i2,j1}+p_{i2,j3}=\Phi_2 \\
p_{i3,j1}+p_{i3,j2}=\Phi_3 \\
\end{cases}
\end{equation}
The system of equations is clearly under defined. Hence, even for the scenario with $V_1=3$ we have three parameters (one per equation). For example, we could set as parameters $p_{i1,j3}$, $p_{i2,j3}$, and $p_{i3,j1}$. Since the values of all the entries are defined between $0$ and $1$, each parameter is bounded by the following conditions: $0\le p_{i1,j3}\le \Phi_1$, $0\le p_{i2,j3}\le \Phi_2$, and $0\le p_{i3,j1}\le \Phi_3$. It is important to stress how once we have picked these three parameters, the matrix $R_{i\alpha,j\beta}$ is fully defined as well as the matrix $R_{j\beta,i\alpha}$ which is the transpose. Furthermore, we have three parameters for each $i<j$, since the entries are in principle function of the age groups. However, for simplicity we set the same three free parameters across all age combinations.

Let us consider now the case in which $V_1=4$. In this case, $W=V_1^2-1=15$. As before, $2V_1-1$ parameters are inputs (three $P_\alpha$ and four $q_\alpha$). In this case, $W>2V_1-1$. Hence, the system of $V_1$ differential equations in $V_1(V_1-1)$ (off-diagonal) elements required $W-2V_1+1=8$ free parameters.

\subsection*{Random mixing regime}

By setting $P_\alpha=q_\alpha=N_\alpha/N$ (where $N_\alpha= \sum_i N_{i\alpha}$) the model drastically simplifies and the contacts division of the total contact $R_{ij}$ between age groups $i$ and $j$ across $\alpha$ and $\beta$ equal the product of the fraction of these two groups (independently of age). This is the random mixing regime, indeed, if contacts are random across groups the expectation value of this process leads to the aforementioned product. To provide a concrete example, let us consider a simple case where we have only one SES dimension and $V_1=3$. As described above, for all $i<j$ in these settings $W=8$ and
\begin{equation}
    p_{i\alpha, j\beta}=R_{ij}\left(
\begin{matrix}
q_1 P_1 & p_{i1,j2}  & p_{i1,j3} \\
p_{i2,j1} & q_2 P_2 & p_{i2,j3} \\
p_{i3,j1}& p_{i3,j2}  & q_3 P_3 \\
\end{matrix}
\right)
\end{equation}
Now we can set the parameters in the diagonal equal to the fraction of the population of each group:
\begin{equation}
    p_{i\alpha,j\beta}=R_{ij}\left(
\begin{matrix}
\frac{N_1}{N} \frac{N_1}{N} & p_{i1,j2}  & p_{i1,j3} \\
p_{i2,j1} &\frac{N_2}{N} \frac{N_2}{N} & p_{i2,j3} \\
p_{i3,j1}& p_{i3,j2}  &\frac{N_3}{N} \frac{N_3}{N} \\
\end{matrix}
\right)
\end{equation}
The off-diagonal values are defined similarly
\begin{equation}
    p_{i\alpha, j\beta}=R_{ij}\left(
\begin{matrix}
\frac{N_1^2}{N^2} & \frac{N_1N_2}{N^2}  & \frac{N_1N_3}{N^2} \\
\frac{N_1N_2}{N^2} &\frac{N_2^2}{N^2} & \frac{N_2N_3}{N^2} \\
\frac{N_1N_3}{N^2}& \frac{N_2N_3}{N^2}  &\frac{N_3^2}{N^2} \\
\end{matrix}
\right)
\end{equation}
The assumption that the division of the $R_{ij}$ contacts for SES depends only on the total fraction of individuals in SES induces symmetry in these matrices even for the case $i<j$. We note how, independently on the number of SES groups, by assigning contacts at random proportionally to the populations allows to define matrices without the need to set free parameters. It is easy to show how by setting $P_\alpha=q_\alpha=N_\alpha/N$, the matrices are defined in such a way that the on-diagonal elements have a $P_\alpha q_\alpha$ fraction of contacts while the off-diagonal elements in each row sums to $P_\alpha(1-q_\alpha)=\Phi_\alpha$

Finally, we note how another way to define an homogeneous mixing scenario could be done considering $R_{i\alpha,j \beta}=R_{ij}\frac{N_{i\alpha}}{N_i}\frac{N_{j\alpha}}{N_j}$. In this case, we still split the total contacts considering the fraction of individuals in each age and SES group. This model does not, in general, lead to the symmetry of the matrices for $i<j$. However, this case reduces to the previous assuming that $\frac{N_{i\alpha}}{N_i}=\frac{N_{\alpha}}{N}$ for $\forall i$, hence when the fraction of individuals in each SES is independent of age.

\subsection*{Epidemic models with generalized contact matrices}

In this section, we provide the detailed description of epidemic models featuring generalized contact matrices. As done above, let us consider a disease whose natural history can be described with a Susceptible-Exposed-Infected-Recovered model~\cite{keeling08}. By considering the population sliced in $m+1$ dimensions the epidemic dynamics are encoded in the following set of differential equations:
\begin{eqnarray}
d_t S_{\mathbf{a}}(t) &=& -\Lambda_\mathbf{a} (t) S_\mathbf{a}(t), \nonumber \\
d_t E_\mathbf{a}(t) &=& \Lambda_\mathbf{a}(t) S_\mathbf{a}(t) -\Psi E_\mathbf{a}(t), \nonumber \\
% d_t E_\mathbf{a}(t) &=& \Lambda S_\mathbf{a}(t) -\Psi E_\mathbf{a}(t), \nonumber \\
d_t I_\mathbf{a}(t) &=& \Psi E_\mathbf{a}(t)-\Gamma I_\mathbf{a} (t),\nonumber \\
d_t R_\mathbf{a}(t) &=& \Gamma I_\mathbf{a}(t).
\end{eqnarray}
where $\mathbf{a}=(i,\alpha, \ldots, \gamma)$ is the index vector describing the membership of individuals in the $m+1$ groups. In other words, the unfolding of the disease is captured by $4\times K \prod_{p=1}^{m}V_p$ compartments. The force of infection, defined as the per-capita rate at which susceptible are exposed to the disease, can be written as:
\begin{equation}
\label{force_G}
    \Lambda_\mathbf{a}(t)= \Phi \sum_{_\mathbf{b}}G_{\mathbf{a},\mathbf{b}}\frac{I_{\mathbf{b}}(t)}{N_\mathbf{b}}
\end{equation}
where $\Phi$ is the trasmissibility of the disease and the temporal dependence is induced by the variation in the number of infected across age brackets. 

\subsection*{Derivation of $R_0$}

The derivation of the basic reproductive number is done following the next generation matrix approach~\cite{blackwood2018introduction}. 
As done for standard contact matrices above, the first step is to focus the attention only on the compartments that describe any stages of the infection: $E_\mathbf{a}$ and $I_\mathbf{a}$. It is convenient to re-write the differential equations for these compartment as 
\begin{equation}
    d_t \mathbf{x}=f(\mathbf{x})-w(\mathbf{x})
\end{equation}
where $\mathbf{x} \in [E_{\mathbf{1}}, \ldots E_{\mathbf{T}}, I_{\mathbf{1}}, \ldots I_{\mathbf{T}}]$, $f(\mathbf{x})$ encodes all terms that lead to new infections and $w(\mathbf{x})$ all other transitions in and out of the compartments. Furthermore, we denote with $\mathbf{a}$ ($a \in [1,T]$ and $T=K\prod_{p=1}^{m}V_p$) the index vectors. In other words, $\mathbf{1}$ describes the index vector $(1,1,\ldots, 1)$ of size $m+1$, $\mathbf{2}$ the index vector $(1,1,\ldots, 2)$, and $\mathbf{T}$ instead $(K,V_1,\ldots, V_m)$. Using this notation, we map the generalized problem to the same structure described above for only one dimension (i.e., age). Indeed, we can write, once again dropping the time dependence for convenience,
\begin{equation}
   d_t \mathbf{x}= \begin{bmatrix}
d_t E_\mathbf{1} \\
\vdots \\
d_t E_\mathbf{T} \\
d_t I_\mathbf{1} \\
\vdots \\
d_t I_\mathbf{T} \\
\end{bmatrix}= f(\mathbf{x})-w(\mathbf{x})=\begin{bmatrix}
\Lambda_\mathbf{1} S_\mathbf{1} \\
\vdots \\
\Lambda_\mathbf{T} S_\mathbf{T}\ \\
0 \\
\vdots \\
0 \\
\end{bmatrix}-\begin{bmatrix}
\Psi E_\mathbf{1} \\
\vdots \\
\Psi E_\mathbf{T} \\
\Gamma I_\mathbf{1}-\Psi E_\mathbf{1} \\
\vdots \\
\Gamma I_\mathbf{T}-\Psi E_\mathbf{T}\\
\end{bmatrix}
\end{equation}
Which is has exactly the same structure of the analogous described above, but where $i \rightarrow \mathbf{a}$ and $\mathbf{a}=(i, \alpha_1, \ldots, \alpha_m)$. As shown above, the expression of $R_0$ is linked to the early epidemic dynamics which can be linearized by calculating the Jacobian at the disease free equilibrium (DFE) $(S^*_{\mathbf{j}},E^*_{\mathbf{j}},I^*_{\mathbf{j}},R^*_{\mathbf{j}})=(N_{\mathbf{j}},0,0,0)$ for all age groups. Using the Jacobian matrix $\mathbf{J}$ (which has size $2K\prod_{p=1}^{m}V_p\times 2K \prod_{p=1}^{m}V_p$) at the DFE we can write 
\begin{equation}
 d_t \mathbf{x}=\mathbf{J}\mathbf{x}   
\end{equation}
This expression can be conveniently factorized as $d_t \mathbf{x}=(\mathbf{F}-\mathbf{W})\mathbf{x}$ where the matrix $\mathbf{F}$ is Jacobian applied to $f(\mathbf{x})$ and similarly the matrix $\mathbf{W}$ is the Jacobian applied to $w(\mathbf{x})$. In particular we have:
\begin{equation}
\footnotesize
    \mathbf{F}=\begin{bmatrix}
\partial_{E_\mathbf{1}}(\Lambda_\mathbf{1} N_\mathbf{1}) & \ldots  & \partial_{E_\mathbf{T}}(\Lambda_\mathbf{1} N_\mathbf{1}) &  \partial_{I_\mathbf{1}}(\Lambda_\mathbf{1} N_\mathbf{1}) & \ldots & \partial_{I_\mathbf{T}}(\Lambda_\mathbf{1} N_\mathbf{1})\\
\partial_{E_\mathbf{1}}(\Lambda_\mathbf{2} N_\mathbf{2}) & \ldots  & \partial_{E_\mathbf{T}}(\Lambda_\mathbf{2} N_\mathbf{2}) &  \partial_{I_\mathbf{1}}(\Lambda_\mathbf{2} N_\mathbf{2}) & \ldots & \partial_{I_\mathbf{T}}(\Lambda_\mathbf{2} N_\mathbf{2}) \\
\vdots & \vdots & \vdots & \vdots & \vdots & \vdots\\
\partial_{E_\mathbf{1}}(\Lambda_\mathbf{T} N_\mathbf{T}) & \ldots  & \partial_{E_\mathbf{T}}(\Lambda_\mathbf{T} N_\mathbf{T}) &  \partial_{I_\mathbf{1}}(\Lambda_\mathbf{T} N_\mathbf{T}) & \ldots & \partial_{I_\mathbf{T}}(\Lambda_\mathbf{T} N_\mathbf{T})\\
0  & \ldots  & 0 & 0 & \ldots  & 0 & \\
0  & \ldots  & 0 & 0 & \ldots  & 0 &\\
\vdots  & \vdots & \vdots & \vdots & \vdots & \vdots\\
0 & \ldots  & 0 & 0 & \ldots  & 0 & \\
\end{bmatrix}=\begin{bmatrix}
0 & \ldots  & 0 &  \partial_{I_\mathbf{1}}(\Lambda_\mathbf{1} N_\mathbf{1}) & \ldots & \partial_{I_\mathbf{T}}(\Lambda_\mathbf{1} N_\mathbf{1})\\
0 & \ldots  & 0 &  \partial_{I_\mathbf{1}}(\Lambda_\mathbf{2} N_\mathbf{2}) & \ldots & \partial_{I_\mathbf{T}}(\Lambda_\mathbf{2} N_\mathbf{2}) \\
\vdots & \vdots & \vdots & \vdots & \vdots & \vdots\\
0 & \ldots  & 0 &  \partial_{I_\mathbf{1}}(\Lambda_\mathbf{T} N_\mathbf{T}) & \ldots & \partial_{I_\mathbf{T}}(\Lambda_\mathbf{T} N_\mathbf{T})\\
0  & \ldots  & 0 & 0 & \ldots  & 0 & \\
0  & \ldots  & 0 & 0 & \ldots  & 0 &\\
\vdots  & \vdots & \vdots & \vdots & \vdots & \vdots\\
0 & \ldots  & 0 & 0 & \ldots  & 0 & \\
\end{bmatrix}
\end{equation}
The components of each gradient are computed at the DFE, i.e., $S_\mathbf{j} \rightarrow N_\mathbf{j}$. By looking at Eq.~\ref{force_G} we note how each partial derivative in $\mathbf{j}$ selects the $j$-th index vector:
\begin{equation}
    \mathbf{F}=\Phi \begin{bmatrix}
0 & \ldots  & 0 &  \frac{G_{\mathbf{1,1}}}{N_\mathbf{1}}N_\mathbf{1} & \ldots & \frac{G_{\mathbf{1,T}}}{N_\mathbf{T}}N_\mathbf{1}\\
0 & \ldots  & 0 &  \frac{G_{\mathbf{2,1}}}{N_\mathbf{1}}N_\mathbf{2} & \ldots & \frac{G_{\mathbf{2,T}}}{N_\mathbf{T}}N_\mathbf{2}\\
\vdots & \vdots & \vdots & \vdots & \vdots & \vdots\\
0 & \ldots  & 0 &  \frac{G_{\mathbf{K,1}}}{N_\mathbf{1}}N_\mathbf{T} & \ldots & \frac{G_{\mathbf{T,T}}}{N_\mathbf{T}}N_\mathbf{T}\\
0 &\ldots  & 0 & 0 & \ldots  & 0 & \\
0 & \ldots  & 0 & 0 & \ldots  & 0 &\\
\vdots & \vdots & \vdots & \vdots & \vdots & \vdots\\
0 & \ldots  & 0 & 0 & \ldots  & 0 & \\
\end{bmatrix}=\Phi \begin{bmatrix}
\mathbf{0} & \mathbf{\tilde{G}} \\
\mathbf{0} &\mathbf{0} 
\end{bmatrix}
\end{equation}
where we denoted with $\mathbf{0}$ $K\prod_{p=1}^{m}V_p\times K\prod_{p=1}^{m}V_p$ blocks of zeros and the generic entry of $\mathbf{\tilde{G}}$ is $\tilde{G}_{\mathbf{a,b}}=\frac{G_{\mathbf{a,b}}}{N_\mathbf{b}}N_\mathbf{a}$. Using the same method we can easily write the expression of the matrix $\mathbf{V_1}$ as:
\begin{equation}
    \mathbf{W}=\begin{bmatrix}
\Psi & \ldots  & 0 &  0 & \ldots  & 0 \\
\vdots & \vdots & \vdots & \vdots & \vdots & \vdots & \vdots & \vdots\\
0 & \ldots  & \Psi & 0 & \ldots  & 0 \\
-\Psi & \ldots  & 0 & \Gamma & \ldots  & 0 & \\
\vdots & \vdots & \vdots & \vdots & \vdots & \vdots & \vdots & \vdots\\
0 & \ldots  & -\Psi & 0 & \ldots  & \Gamma & \\
\end{bmatrix}= \begin{bmatrix}
\Psi\mathbf{I} & \mathbf{0} \\
-\Psi\mathbf{I} & \Gamma\mathbf{I} 
\end{bmatrix}
\end{equation}
where $\mathbf{I}$ are diagonal matrices of size $K\prod_{p=1}^{m}V_p \times K\prod_{p=1}^{m}V_p$. The expression for $R_0$ is linked to the two matrices as $R_0=\rho(\mathbf{FW}^{-1})$, where $\rho$ denotes the spectral radius. It is easy to show how
\begin{equation}
    \mathbf{FW}^{-1}=\Phi \begin{bmatrix}
\mathbf{0} & \mathbf{\tilde{G}} \\
\mathbf{0} &\mathbf{0} 
\end{bmatrix}\begin{bmatrix}
\frac{1}{\Psi}\mathbf{I} & \mathbf{0} \\
\frac{1}{\Gamma}\mathbf{I} & \frac{1}{\Gamma}\mathbf{I} 
\end{bmatrix}=\frac{\Phi}{\Gamma} \begin{bmatrix}
\mathbf{\tilde{G}} & \mathbf{\tilde{G}} \\
\mathbf{0} &\mathbf{0} 
\end{bmatrix}
\end{equation}
hence, we can finally write
\begin{equation}
    R_0=\frac{\Phi}{\Gamma} \rho(\mathbf{\tilde{G}})
\end{equation}
The expression has the same general form of the one derived for standard contact matrices stratified just for age. However, the entries of the matrix are different and those encode the multiple dimensions.
Contact matrices are often stratified also for the context (location) where they take place~\cite{Mistry2021}. The entries $G_{\mathbf{a;b}}$ are then expressed as
\begin{equation}
    G_{\mathbf{a,b}}=\sum_{l}\omega_l G_{\mathbf{a,b}}^{(l)}
\end{equation}
where $\omega_s$ are weights capture possible heterogeneities in the relevance of contacts in each context in the transmission of the disease. In this formulation, the inclusion of the different contexts does not change the expression of $R_0$ as the $\omega_l$ are assumed to be homogeneous across age-brackets.

\subsection*{On the spectral radius of generalized contact matrices}

We provide some formal relations between the spectral radius of the generalized contacts matrices and the classical ones. We denote $\tilde{\mathbf{C}}$ in matrix form as $\tilde{\mathbf{C}} = \mathbf{R}\mathbf{N}^{-1}$, where $\mathbf{N}$ is diagonal and stores the number of individuals in each age group and $\mathbf{R}$ is symmetric and contains the number of contacts between pairs. We can similarly denote the generalized contact matrix $\mathbf{\tilde{G}} = \mathbf{R_G}\mathbf{N_G}^{-1}$, where $\mathbf{N_G}$ is now the diagonal matrix of the generalized populations sizes. As we stated already, these two contact matrices have clearly different sizes that we denote as $\tilde{\mathbf{C}} \in \mathbb{R}^{K\times K}$ and $\tilde{\mathbf{G}} \in \mathbb{R}^{KV_1 \times KV_1}$. To move from $\tilde{\mathbf{C}}$ to $\tilde{\mathbf{G}}$ we thus need an ``inflation'' process that turns every element of $\mathbf{\tilde{C}}$ into a matrix of size $V\times V$. We define three inflation techniques and provide a theorem relating the spectral radius of $\mathbf{\tilde{C}}$ and $\mathbf{\tilde{G}}$ in these three cases.

\medskip

As a fundamental remark, we here only treat inflation strategies that go from the indexing $(i,j)$ to $(i\alpha, j\beta)$,  i.e., that add only one dimension. This can be done without loss of generality, because if multiple dimensions exist, one can simply add them iteratively one by one.

\subsection*{Theorem enunciation}

\begin{definition}[Homogeneous mixing inflation]
	\label{def:hom}
	The homogeneous mixing inflation creates $\mathbf{\tilde{G}}$ from $\mathbf{\tilde{C}}$ assuming nothing is known about the newly added dimension. In this case we get $\mathbf{\tilde{G}}_{\rm hom} = \mathbf{R}_{\mathbf{G},\rm hom}\mathbf{N}_{\mathbf{G}, {\rm hom}}^{-1}$, where
	\begin{align*}
	(\mathbf{R}_{{\mathbf G},\rm hom})_{i\alpha, j\beta} &= \frac{\mathbf{R}_{ij}}{V^2}\\
	(\mathbf{N}_{\mathbf{G}, {\rm hom}})_{i\alpha, j\beta} &= \frac{N_i}{V_1}\delta_{ij}\delta_{\alpha\beta}.
	\end{align*}
\end{definition}

The second inflation technique is the random mixing

\begin{definition}[Random mixing inflation]
	\label{def:rdn}
	The random mixing inflation is obtained assuming that the sub-population sizes are known and that the number of contacts is proportional to the product of interacting group sizes. More precisely, $\mathbf{\tilde{G}}_{\rm rdn} = \mathbf{R}_{{\mathbf G}, \rm rdn}\mathbf{N}_{\mathbf{G}, {\rm rdn}}^{-1}$, where
	\begin{align*}
	(\mathbf{R}_{{\mathbf G}, \rm rdn})_{i\alpha, j\beta} &= \mathbf{R}_{ij}\frac{N_\alpha N_\beta}{N^2}\\
	(\mathbf{N}_{\mathbf{G}, {\rm rdn}})_{i\alpha, j\beta} &= \frac{N_{i}N_{\alpha}}{N}\delta_{ij}\delta_{\alpha \beta},
	\end{align*}
	where $N_{\alpha}$ is the size of group $\alpha$ and $N$ is the total number of individuals.
\end{definition}

We finally consider a third inflation strategy that is general and model-free. This technique includes all possible ways to move from $\mathbf{\tilde{C}}$ to $\mathbf{\tilde{G}}$, including the two we just described, the assortative mixing model and the real matrix $\mathbf{\tilde{G}}$.

\begin{definition}[Arbitrary inflation]
	\label{def:arb}
	An arbitrary inflation is obtained without assuming any contact pattern and just requiring to preserve the population sizes and the number of contacts. More precisely, we let $\mathbf{\tilde{G}}_{\rm a} = \mathbf{R}_{{\mathbf G}, \rm a}\mathbf{N}_{\mathbf{G}, {\rm a}}^{-1}$, where
	\begin{align*}
	\sum_{\alpha \beta}(\mathbf{R}_{{\mathbf G}, \rm a})_{i\alpha, j\beta} &= \mathbf{R}_{ij}\\
	\sum_{\alpha\beta}(\mathbf{N}_{\mathbf{G}, {\rm a}})_{i\alpha, j\beta} &= \mathbf{N}_{ij}.
	\end{align*}
\end{definition}

We now provide our main result relating the spectral radius of $\mathbf{\tilde{G}}$ with that of $\mathbf{\tilde{C}}$ in these three cases.

\begin{theorem}[Spectral radius of generalized contact matrices]
	Consider a contact matrix $\mathbf{\tilde{C}}$ and its generalized version $\mathbf{\tilde{G}}$ obtained with arbitrary inflation as per Definition~\ref{def:arb}. Then,
	\begin{align*}
	\rho(\mathbf{\tilde{G}}) \geq \rho(\mathbf{\tilde{C}}).
	\end{align*}
	
	Moreover, the strict equality is obtained for the homogeneous and random mixing (Definitions~\ref{def:hom}, \ref{def:rdn}).
\end{theorem}

\subsection*{Proof}

We now proceed with the proof. For simplicity, we hereby summarize the strategy. First we show that in the two cases in which the equality holds, the leading eigenvector can be easily derived analytically hence its related eigenvalue. Then, exploiting the fact that both $\mathbf{\tilde{C}}, \mathbf{\tilde{G}}$ have the same eigenvalues of a symmetric matrix, we compute the Rayleigh quotient of $\mathbf{\tilde{G}}_{\rm a}$ for a guess vector and prove that it equals $\rho(\mathbf{\tilde{G}}_{\rm hom})$ thus concluding the proof.

\subsubsection*{Definition of the symmetrized contact matrices}

To ease the calculation, we adopt a symmetrized version of $\mathbf{\tilde{C}}, \mathbf{\tilde{G}}$ that share the same eigenvalues \cite[Corollary 111.4.6]{bhatia2013matrix}:

\begin{align*}
\mathbf{\tilde{C}}^{\rm sym} &= \mathbf{N}^{-1/2}\mathbf{R}\mathbf{N}^{-1/2}\\
\mathbf{\tilde{G}}^{\rm sym} &= \mathbf{N_G}^{-1/2}\mathbf{R_G}\mathbf{N_G}^{-1/2}.
\end{align*}

Since the symmetric matrices are similar to their non-symmetric counterpart, we can write

\begin{align*}
\rho(\mathbf{\tilde{G}}^{\rm sym}) \geq \rho(\mathbf{\tilde{C}}^{\rm sym}) \iff \rho(\mathbf{\tilde{G}}) \geq \rho(\mathbf{\tilde{C}}).
\end{align*}

For simplicity and uniformity we now only work with the symmetric matrices.

\subsubsection*{Spectral radius for the homogeneous inflation}

Given Definition~\ref{def:hom}, we can write the $(i\alpha, j\beta)$ element of $\mathbf{\tilde{G}}_{\rm hom}^{\rm sym}$ as

\begin{align*}
\left(\mathbf{\tilde{G}}^{\rm sym}_{\rm hom}\right)_{i\alpha,j\beta} = \frac{\mathbf{\tilde{C}}_{ij}^{\rm sym}}{V_1}.
\end{align*}

Let $\mathbf{x} \in \mathbb{R}^K$ be the leading eigenvector of $\mathbf{\tilde{C}}^{\rm sym}$ and let $\rho$ be its associated eigenvalue. Then we define $\mathbf{y} \in \mathbb{R}^{KV_1}$ so that $\mathbf{y}_{i\alpha} = \mathbf{x}_i$. We now show that $\mathbf{y}$ is the (un-normalized) leading eigenvector of $\mathbf{\tilde{G}}_{\rm hom}^{\rm sym}$ with eigenvalue $\rho$, in fact

\begin{align*}
\left(\mathbf{\tilde{G}}_{\rm hom}^{\rm sym}\mathbf{y}\right)_{i\alpha} = \sum_{j\beta} \left(\mathbf{\tilde{G}}_{\rm hom}^{\rm sym}\right)_{i\alpha, j\beta}\mathbf{y}_{j\beta} = \frac{1}{V_1}\sum_{j\beta} \mathbf{\tilde{C}}^{\rm sym}_{ij}\mathbf{x}_j = \sum_{j} \mathbf{\tilde{C}}^{\rm sym}_{ij}\mathbf{x}_j = \left(\mathbf{\tilde{C}}^{\rm sym}_{ij}\mathbf{x}\right)_i = \rho \mathbf{x}_i = \rho \mathbf{y}_{i\alpha}.
\end{align*}

Note that, due to Perron-Frobenius theorem, $\rho$ must be the largest eigenvalue of $\mathbf{\tilde{G}}_{\rm hom}^{\rm sym}$.

\subsubsection*{Spectral radius under the random inflation}

Given Definition~\ref{def:rdn} we may write the $(i\alpha, j\beta)$ component of $\mathbf{\tilde{G}}_{\rm rdn}^{\rm sym}$ as

\begin{align*}
\left(\mathbf{\tilde{G}}_{\rm rdn}^{\rm sym}\right)_{i\alpha, j\beta} = \sqrt{\frac{N_\alpha N_\beta}{N_iN_j}}\cdot\frac{\mathbf{R}_{ij}}{N}.
\end{align*}

Once again, we denote with $\mathbf{x} \in \mathbb{R}^K$ the leading eigenvector of $\mathbf{\tilde{C}}^{\rm sym}$ and with $\rho$ its associated eigenvalue. We define $\mathbf{y} \in \mathbb{R}^{KV_1}$ so that $\mathbf{y}_{i\alpha} = \mathbf{x}_i\sqrt{N_\alpha}$ and we show that it is the (un-normalized) leading eigenvector of $\mathbf{\tilde{G}}_{\rm rdn}^{\rm sym}$ with eigenvalue $\rho$.

\begin{align*}
\left(\mathbf{\tilde{G}}_{\rm rdn}^{\rm sym}\mathbf{y}\right)_{i\alpha} &= \sum_{j\beta}\left(\mathbf{\tilde{G}}_{\rm rdn}^{\rm sym}\right)_{i\alpha,j\beta}\mathbf{y}_{j\beta} = \sum_{j\beta} \sqrt{\frac{N_\alpha N_\beta}{N_iN_j}}\cdot \frac{\mathbf{R}_{ij}}{N} \mathbf{x}_j\sqrt{N_\beta} = \frac{\sqrt{N_\alpha}}{N}\sum_\beta N_\beta \sum_j \mathbf{\tilde{C}}^{\rm sym}_{ij}\mathbf{x}_j\\
&= \rho\sqrt{N_\alpha}\mathbf{x}_i = \rho \mathbf{y}_{i\alpha}.
\end{align*}

Once again, this is the largest eigenvalue due to Perron-Frobenius theorem.

\subsubsection*{Spectral radius for arbitrary inflation}

We here provide an inequality for the spectral radius of $\mathbf{\tilde{G}}_{\rm a}^{\rm sym}$, exploiting a known property that the largest eigenvalue of a Hermitian matrix is the maximum of the Rayleigh quotient. By definition, we can write

\begin{align*}
\left(\mathbf{\tilde{G}}_{\rm a}^{\rm sym}\right) = \frac{\left(\mathbf{R}_{{\mathbf G}, {\rm a}}\right)_{i\alpha, j\beta}}{\sqrt{\left(N_{{\mathbf G}, {\rm a}}\right)_{i\alpha}\left(N_{{\mathbf G}, {\rm a}}\right)_{j\beta}}}.
\end{align*}

Consider the vector $\mathbf{y} \in \mathbb{R}^{K\times V_1}$ so that $\mathbf{y}_{i\alpha} = \sqrt{\frac{\left(N_{\mathbf{G}, {\rm a}}\right)_{i\alpha}}{N_i}} \mathbf{x}_i$, where $\mathbf{x}$ is the usual vector. We can write

\begin{align*}
\rho\left(\mathbf{\tilde{G}}_{\rm a}^{\rm sym}\right) &= \underset{\mathbf{z} \in \mathbb{R}^{VK} \neq \mathbf{0}}{\rm max}\frac{\mathbf{z}^T\mathbf{\tilde{G}}_{\rm a}^{\rm sym}\mathbf{z}}{\mathbf{z}^T\mathbf{z}} \geq \frac{\mathbf{y}^T\mathbf{\tilde{G}}_{\rm a}^{\rm sym}\mathbf{y}}{\mathbf{y}^T\mathbf{y}} = \frac{\sum_{i\alpha}\sum_{j\beta}\sqrt{\frac{\left(N_{\mathbf{G}, {\rm a}}\right)_{i\alpha}\left(N_{\mathbf{G}, {\rm a}}\right)_{j\beta}}{N_iN_j}}\frac{\left(\mathbf{R}_{{\mathbf G}, {\rm a}}\right)_{i\alpha, j\beta}}{\sqrt{\left(N_{{\mathbf G}, {\rm a}}\right)_{i\alpha}\left(N_{{\mathbf G}, {\rm a}}\right)_{j\beta}}}\mathbf{x}_i\mathbf{x}_j}{\sum_{i\alpha}\frac{\left(N_{\mathbf{G}, {\rm a}}\right)_{i\alpha}}{N_i} \mathbf{x}_i^2}\\
&= \frac{\sum_{ij}\frac{\mathbf{x}_i\mathbf{x}_j}{\sqrt{N_iN_j}}\sum_{\alpha\beta}\left(\mathbf{R}_{\mathbf{G}, {\rm a}}\right)_{i\alpha,j\beta}}{\sum_i \frac{\mathbf{x}_i^2}{N_i}\sum_\alpha \left(N_{\mathbf{G}, {\rm a}}\right)_{i\alpha}} = \frac{\sum_{ij}\frac{\mathbf{x}_i\mathbf{x}_j}{\sqrt{N_iN_j}}\mathbf{R}_{ij}}{\sum_i \frac{\mathbf{x}_i^2}{N_i}N_i} = \frac{\sum_{ij}\mathbf{\tilde{C}}^{\rm sym}_{ij}\mathbf{x}_i\mathbf{x}_j}{\Vert \mathbf{x}\Vert^2} = \rho,
\end{align*}

thus $\rho\left(\mathbf{\tilde{G}}_{\rm a}^{\rm sym}\right) \geq \rho$, concluding the proof.

%---- begin adriana
%\subsection*{Epidemic simulations}

%To investigate the effect of the generalized contact matrices $G_{\mathbf{a},\mathbf{b}}$ on infection transmission dynamics, we developed a stochastic, discrete-time, compartmental model where the transitions among compartments are simulated through chain binomial processes. In particular, at time step $t$ the number of individuals in group $\mathbf{a}$ and compartment $X$ transiting to compartment $Y$ is sampled from $PrBin(X\mathbf{a}(t),p_{X\mathbf{a}\xrightarrow{}Y\mathbf{a}}(t))$,where $p_{X\mathbf{a}\xrightarrow{}Y\mathbf{a}}(t)$ is the transition probability.

%As an example, let's consider the number of individuals in group $\mathbf{a}$ and compartment $S$ that at time $t$ become infected transiting in compartment $I$. Thus, the number of individuals in $S_{\mathbf{a}}(t)$ getting infected are extracted from a $PrBin(S_{\mathbf{a}}(t), \Lambda_{\mathbf{a}}(t)$ where $ \Lambda_{\mathbf{a}}(t)$ is the \textit{force of infection} as defined in eq.\ref{force_G}.

\subsection*{Data}

\subsubsection*{The MASZK questionnaire} In this study we used data coming from the MASZK survey study \cite{karsai2020hungary,koltai2022reconstructing}, a large data collection effort on social mixing patterns made during the COVID-19 pandemic, conducted in Hungary from April 2020 to July 2022. The study involved $26$ monthly cross-sectional anonymous phone surveys using Computer Assisted Telephone Interview (CATI) methodology, with a nationally representative sample of 1000 participants each month. The recorded population was representative in terms
of gender, age, education level and type of settlement; sampling errors were further corrected by post-stratification weights. The data collection adhered to European and Hungarian privacy regulations, approved by the Hungarian National Authority for Data Protection and Freedom of Information \cite{naih}, as well as the Health Science Council Scientific and Research Ethics Committee (resolution number IV/3073-1/2021/EKU).

Relevant to this study, the questionnaires recorded information about the \emph{proxy social contacts}, defined as interactions where the respondent and a peer stayed within 2 meters for more than 15 minutes~\cite{ProxyContactDef}, at least one of them without wearing a mask. Approximate contact numbers were recorded between the respondents and their peers from different age groups of 0–4, 5–14, 15–29, 30–44, 45–59, 60–69, 70–79, and 80+. Contact number data about underage children were collected by asking legal guardians to estimate daily contact patterns. Participants during the while data collection were asked to report contacts referring $(i)$ to the previous day (that we use for the analysis in Fig 4 in the main text) and, during the first data collection campaigns $(ii)$ to an average pre-pandemic day (that we use for the analysis in Fig 2 and 3 in the main text). Additionally, in three data collection waves: April 2021, November 2021 and June 2022 contact have been collected in the form of diaries. Namely, participants were asked to list one by one the contacts they had on the previous day by providing some socio-demographic information about the contacts such as their wealth situation. Beyond information on contacts before and during the pandemic, the MASZK dataset provided us with information on \textit{social-demographic characteristics} of participants, such as their \textit{perceived wealth situation}, \textit{gender}, \textit{vaccination status}, etc. 

%On the collected data a multi-step, proportionally stratified, probabilistic sampling procedure was elaborated and implemented by the survey research company using a database that contained both landline and mobile phone numbers. The survey response rate was 49 per cent, which is expressly higher than the average response rate (being between 15-20 per cent) of telephone surveys in Hungary. The sample is representative of the Hungarian population aged 18 or older by gender, age, education and domicile. Sampling errors were corrected using iterative proportional post-stratification weights. After data collection, only the anonymised and hashed data was shared with people involved in the project after signing non-disclosure agreements.

\subsection*{Other contacts data} Beyond to the data on Hungary, we use data on Zimbabwe age-contact patterns \cite{melegaro2013_zimb}. Also in this case, the contacts numbers among age classes have been collected trough surveys. 

\subsection*{Age-contact matrices}

Although theoretically contacts (i.e., raw contacts) are expected to be symmetrical, due to survey limitations and reporting errors they do not achieve perfect reciprocity. Given that our model to build synthetic generalized contact matrices respect the symmetric assumption on the raw contacts, we needed to ensure that also the matrix $\mathbf{R}$ that we fed to these models respects this assumption as well. Thus, all the raw contact matrices used in this study have been corrected for reciprocity \textit{a priori}. This consists in averaging the total number of contacts measured in one direction, from $i$ to $j$, and the reciprocal from $j$ to $i$ (Eq \ref{eq_rec}) \cite{arregui2018projecting}. 

\begin{equation}
    C_{ij} = \frac{C_{ij}^{data}N_i + C_{ji}^{data}N_j}{2N_i}=\frac{R_{ij}}{N_i}
    \label{eq_rec}
\end{equation}

Where, $N_i$ indicates the number of participants in age class $i$ in each study, and $C_{ij}^{data}N_i$ is the total number of contacts reported in the survey from individuals in age class $i$ with individuals in age class $j$.

\begin{figure}[H]
\centering
\includegraphics[scale=0.54]{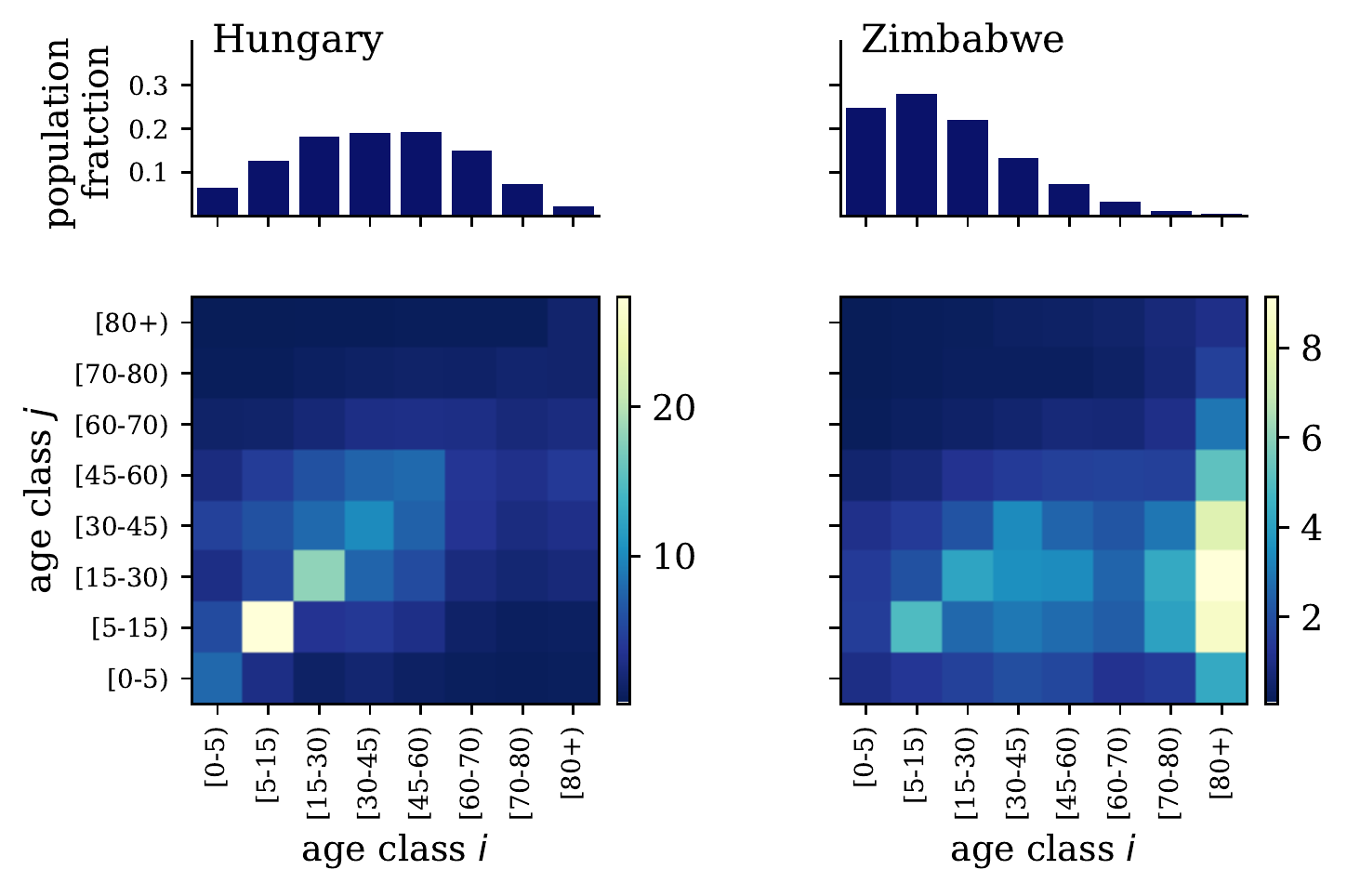}
\caption{($1{st}$ \textit{row})  Population distribution by age ($2{nd}$ \textit{row}) Age-contact matrices refered to the pre-pandemic period for Hungary and Zimbabwe.}
\label{pop_mat}
\end{figure}

\subsection*{Parameters and implementation}
In order to ensure reproducibility of our work, in this section we include a comprehensive list of all the parameters employed to generate the synthetic and real-world generalized contact matrices utilized in the analysis presented in the main text. In addition, we provide all the technical details on the implement of the analysis. 

\subsection*{Figure 2} In Table \ref{tab_fig2} we report the list of parameters used to generate the generalized contact matrices for the analysis presented in Figure 2 in the main text. For the random mixing scenario the only set of parameters of interest is the population distribution. While, for the assortative regime we also report the values of the activity, assortativity and for the free parameters. 

\begin{table}[h!]
\centering
\begin{tabular}{@{}llll@{}}
\toprule
Level & \multicolumn{1}{c}{\textbf{1}} & \multicolumn{1}{c}{\textbf{2}} & \multicolumn{1}{c}{\textbf{3}} \\ \midrule
Population distribution & 35\% & 45\% & 20\% \\
Assortativity ($q_i$) & 60\% & 50\% & 65\% \\
Activity ($P_i$) & 20\% & 40\% & 40\% \\
Free parameters & 0.6 & 0.6 & 0.5 \\ \bottomrule
\end{tabular}
\caption{List of parameters used to generate Figure 2 in the main text. 1,2 and 3 refers to the three categories of dimension two.}
\label{tab_fig2}

\end{table}

\subsection*{Figure 3}  To demonstrate the flexibility of the proposed approach to model NPIs in Figure 3 we analyze three different scenarios: $(i)$ Baseline, $(ii)$ NPI 1 and $(iii)$ NPI 2. 
In particular, in the scenario of NPI 1, we assume that all the individuals reduce their contacts by 35\% at $t^*=50$, while in the scenario of NPI 2, we assume that the introduction of NPI also affects the structure of the matrix. Thus, to model these scenarios, we first reduce by 35\% the age-contact matrix of the total number of contacts such that $R_{ij}^{NPI}= 0.65R_{ij}$. Then, we feed $R_{ij}^{NPI}$ to our model to generate synthetic generalized contact matrices. For the first scenario (NPI 1), the parameters of the model are equal to one of the baseline scenario, while, the parameters of the second scenario (NPI 2) change accordingly to the assumption that we made on the new structure of contacts. In Table \ref{tab_fig3} we report the list of parameters used to generate the generalized contact matrices used for this analysis.

\begin{table}[h!]
\centering
\begin{tabular}{@{}lllllll@{}}
\toprule
\textit{} & \multicolumn{3}{c}{\textbf{Baseline}} & \multicolumn{3}{c}{\textbf{NPI 2}} \\ \midrule
\multicolumn{1}{l|}{Level} & \multicolumn{1}{c}{\textbf{1}} & \multicolumn{1}{c}{\textbf{2}} & \multicolumn{1}{c|}{\textbf{3}} & \multicolumn{1}{c}{\textbf{1}} & \multicolumn{1}{c}{\textbf{2}} & \multicolumn{1}{c}{\textbf{3}} \\ \midrule
\multicolumn{1}{l|}{Population distribution} & 33,3\% & 33,3\% & \multicolumn{1}{l|}{33,3\%} & 33,3\% & 33,3\% & 33,3\% \\
\multicolumn{1}{l|}{Assortativity} & 33,3\% & 50\% & \multicolumn{1}{l|}{60\%} & 50\% & 60\% & 70\% \\
\multicolumn{1}{l|}{Activity ($P_i$)} & 25\% & 45\% & \multicolumn{1}{l|}{30\%} & 37\% & 37\% & 26\% \\
\multicolumn{1}{l|}{Free parameters} & 0.5 & 0.5 & \multicolumn{1}{l|}{0.5} & 0.5\% & 0.4\% & 0.4\% \\ \bottomrule
\end{tabular}
\caption{List of parameters used to generate Figure 3 in the main text, for $(i)$ baseline and, $(ii)$ NPI 2 scenarios. 1,2 and 3 refers to the three categories of dimension two.}
\label{tab_fig3}
\end{table}

\subsection*{Figure 4} We built generalized contact matrices stratified in two dimensions by using real data from the \textit{MASZK} study.

Information on social interactions have been collected in two different ways i) in an aggregate form, such that each participant reported the number of contacts they had with individuals in each of the eight age bracket considered, ii) in a diary in which each participant listed one by one the interactions they had on a given day by reporting some meta information of the \textit{contactee} such as their age and SES. In particular, the average number of contacts of an individual in age class $i$, and SES $\alpha$ with an individual in age class $j$, and SES $\beta$ is $G_{\mathbf{a,b}}$ where $\mathbf{a}=[i,\alpha]$ and $\mathbf{b}=[j,\beta]$. However, the \textit{MASZK} data provided us with diaries information only for the adult population (individuals older than 15 years old). For the children, their average number of contacts is available only in the aggregate form for the eight age brackets considered. Thus, we assigned the average number of contactee to the different SES as follow:
$G_{i\alpha, j \beta} = G_{i \alpha, j} u_{\alpha \beta}$, where $G_{i \alpha, j}$ is the average number of contacts that individual of age group $i$ and SES $\alpha$ has with all the individuals of age group $j$, and $u_{\alpha \beta}$ is a parameter that controls how these contacts are distributed among individuals of different SES.

In particular,to mimic assortativity patterns among SES,the values of $u_{\alpha \beta}$ is are set as follow.

\begin{equation}
   u_{\alpha\beta}=\left(
\begin{matrix}
0.7 & 0.2  & 0.1 \\
0.1 & 0.7  & 0.2 \\
0.1 & 0.2  &0.7  \\
\end{matrix}
\right)
\end{equation}

The average number of contact of these matrices includes the contacts of the family members. However there contacts have been added afterwords to the matrix by assuming that individual belonging to the same family have the same SES.

\subsection*{Impact of generalised contact matrices on epidemic modeling}

\subsubsection*{Zimbabwe}
\begin{figure*}
    \centering
    \includegraphics[scale=0.46]{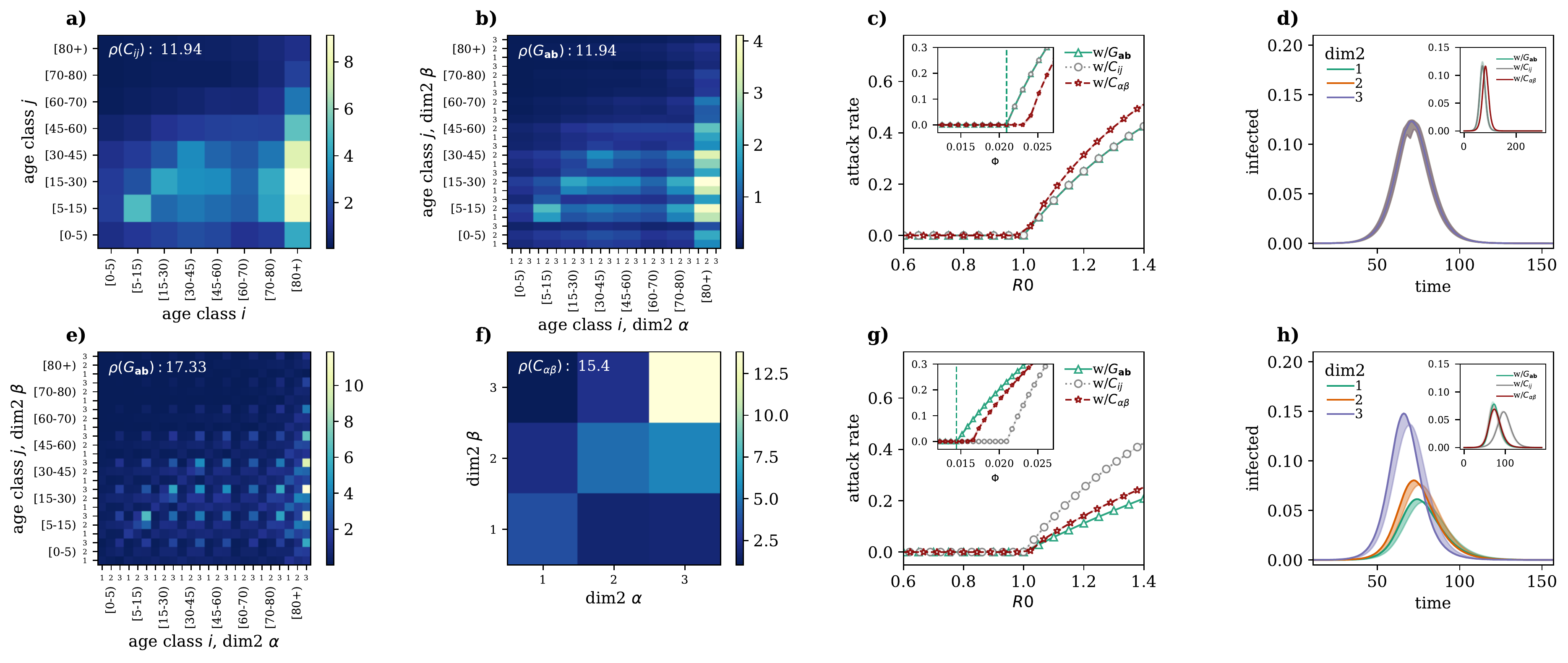}
    \caption{\textbf{Impact of generalized contact matrices on epidemic spreading}. (a) Example of a classic age contact matrix $C_{ij}$ ($8 \times 8$); (b) and (e) Generalized contact matrices with an additional dimension with three levels $G_{\mathbf{a}\mathbf{b}}$ ($24 \times 24$) in the case of random mixing among the second level groups (b) and in the case of assortative mixing and different activity levels among the second dimension (e); (c) attack rate as a function of $R_0$ and disease prevalence (d) as a function of time in the case of random mixing; (g) and (h) as the previous two plots but for the case of assortative mixing with different levels of activity. Results refer to the median of 500 runs. Epidemiological parameters: $\mu=0.25, \epsilon=0.4$, $R_0=2.5$, $I_0= 100$}  
    \label{panel_2_SI}
\end{figure*}

To provide evidence of the robustness of the proposed approach we analyze contact data coming from another country. In particular, we consider the age-contact matrix $\mathbf{C}$ of Zimbabwe \cite{melegaro2013_zimb} (Fig.~\ref{panel_2_SI}-a). We note how the age contact matrix contact is different from the Hungarian one. Indeed, although the age assortative and the inter-generational interaction are still clear, it shows a high activity rate for the 80+.
By applying the same analysis that we did for the Hungarian data we study the impact of generalized contact matrices on epidemic modelling in $(i)$ the random mixing (Fig.~\ref{panel_2_SI}-b) and,  $(ii)$ the assortative mixing (Fig.~\ref{panel_2_SI}-e) regimes. 
As in the main text, we imagine a simple case where the second dimension contains three groups: $1, 2, 3$. We assume that respectively $35\%$, $45\%$, and $20\%$ of the population belong to these three categories across all age groups for simplicity. For the assortative nixing scenario we assume that $60\%$, $50\%$, and $65\%$ of the contacts in the first, second, and third group of the additional category take place within each group. Additionally, we assume some levels of heterogeneity also in the activity of the different groups setting as $20\%$, $40\%$, and $40\%$ the share of contacts of the three groups. In Fig.~\ref{panel_2_SI}-f we show the matrix $C_{\alpha \beta}$ obtained by integrating the latter generalized contact matrix over all age classes.

We use these matrices to study the unfolding of a virus in the population. In Fig.~\ref{panel_2_SI}-c we plot the attack rate (i.e., epidemic size) as a function of $R_0$ and as a function of the transmissibility parameter $\Phi$ (inset) for $1)$ a model fed only with the contact matrix $C_{ij}$ (grey circles), $2)$ a model fed with the generalized contact matrix $G_{\mathbf{a},\mathbf{b}}$ (green triangles), $3)$ a model fed with the matrix $C_{\alpha \beta}$ where contacts are stratified only according to the second dimension (red stars). In Fig.~ \ref{panel_2_SI}-d we show the prevalence of infected overtime of the three dimensions as predicted by the model fed with $G_{\mathbf{a},\mathbf{b}}$  for the three groups. Instead in the inset, we show the the overall prevalence of infected predicted by the three models. We perform the same analysis for the assortative scenario and we show the results respectively in Fig.~ \ref{panel_2_SI}-g and Fig.~\ref{panel_2_SI}-h. All the results are in perfect agreement with the ones discussed in the main text.

\subsubsection*{Sensitivity analysis}
In this section, we perform a sensitivity analysis of the model we propose to build synthetic generalized contact matrices.
In particular, we show how the spectral radius of $G_{\mathbf{a}\mathbf{b}}$ differs from the one computed on the aggregate age contact $C_{ij}$  when we vary some of the parameters of the model.

\begin{figure*}
    \centering
    \includegraphics[width=0.83\textwidth]{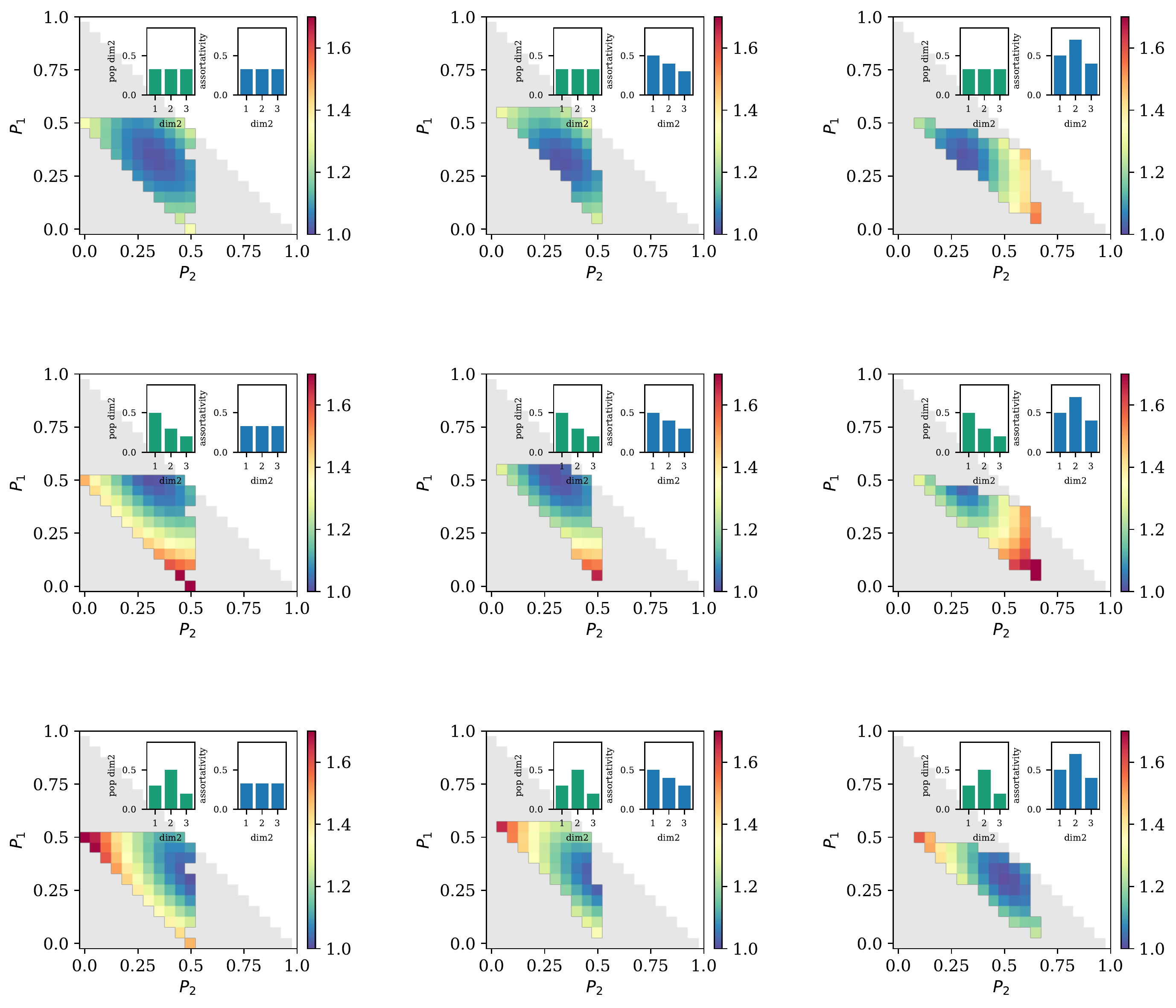}
    \caption{\textbf{Sensitivity analysis}. Variation of the ratio of the $\rho(G_{\mathbf{a}\mathbf{b}})$ over $\rho(C_{ij})$ as function of the activities $P_1, P_2$. Given that $\sum_1^3 P_i =1$ for each $P_1$  and $P_2$,  $P_3 = 1- P_1-P_2$.  In each subplot $G_{\mathbf{a}\mathbf{b}}$ is computed according to our synthetic model using different parameters distribution of population distribution of the second dimension (left inset) and assortativity along the second dimension (right inset). The grey area refers to the non-physical solutions of the model.}
    \label{sensitivity}
\end{figure*}

We create nine different scenarios one for each combination of three sets of values for the population distribution of the second dimension and three sets of parameters for the assortativity. For each of these scenarios in Fig. \ref{sensitivity} we show the ratio of the $\rho(G_{\mathbf{a}\mathbf{b}})$ over $\rho(C_{ij})$ as function of the activity of individuals in dim1 ($P_1$) and, of individuals in dim2 ($P_2$). Given that $\sum_1^3 P_i =1$ for each $P_1$  and $P_2$,  $P_3 = 1-P_1-P_2$, thus each pair of values of $P_1$, $P_2$ defines also the value of $P_3$. 

As shown above, $\rho(G_{\mathbf{a}\mathbf{b}})$ is always grater or equal to $\rho(C_{ij})$ as expected from the theory. From Fig. \ref{sensitivity} we can observe how the parameter sets of the different scenarios are influencing the magnitude of this relation, suggesting that different interaction patterns arising from correlation to an additional dimension effectively impact the estimation of the basic reproductive number. 

\clearpage

\subsubsection*{Testing the model with three dimensions}

\begin{figure*}
    \centering
    \includegraphics[scale=0.46]{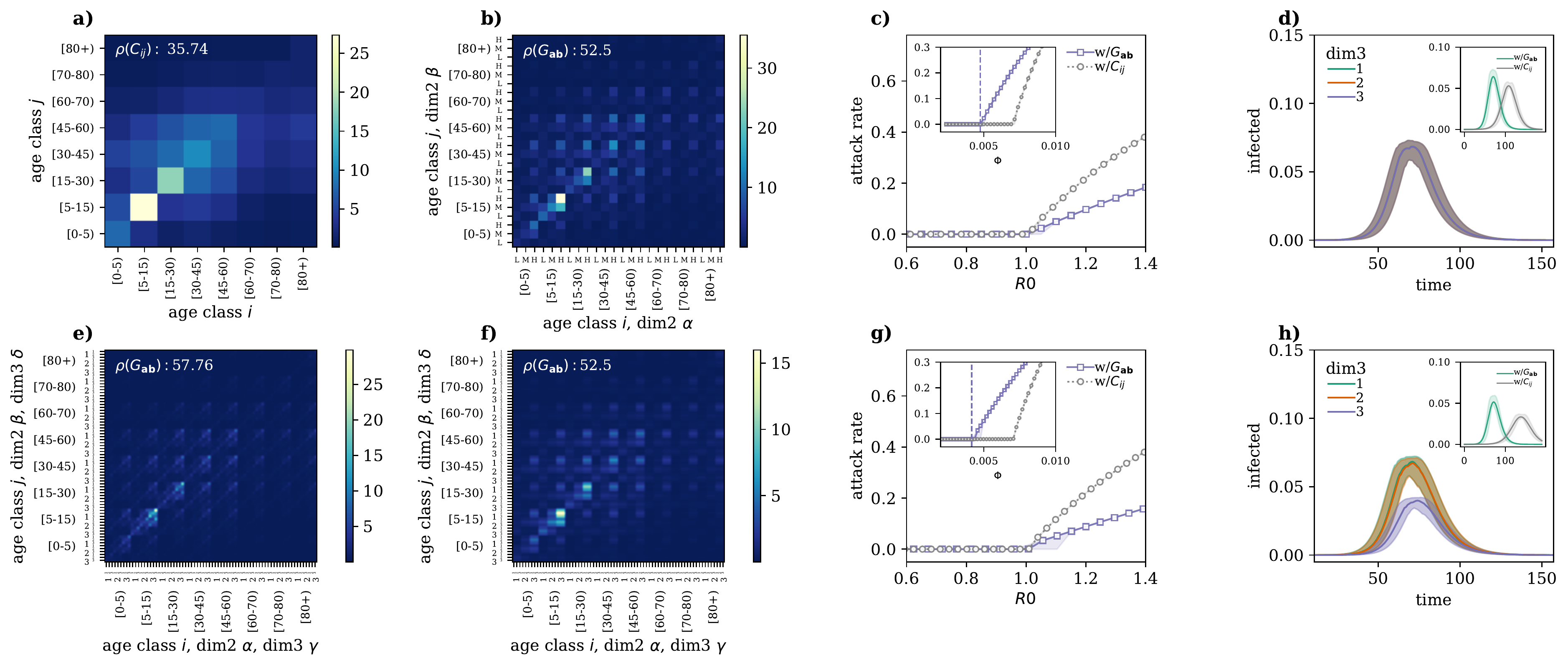}
    \caption{\textbf{Three dimensions generalized contact matrices}. (a) Example of a classic age contact matrix $C_{ij}$ ($8 \times 8$); (b) Generalized contact matrices with an additional dimension with two levels $G_{\mathbf{a}\mathbf{b}}$ ($24 \times 24$) in the case of assortative mixing and different activity levels among the second dimension (e) and (f) Generalized contact matrices with three levels $G_{\mathbf{a}\mathbf{b}}$ ($72 \times 72$)  (e) in the case of assortative mixing and different activity levels among the third dimension and (f) in the case of random mixing among the third dimension (e); (c) attack rate as a function of $R_0$ and disease prevalence (d) as a function of time in the case of random mixing; (g) and (h) as the previous two plots but for the case of assortative mixing with different levels of activity. The results in panel (c),(d),(g),(h) refers to the model with three levels. Results refer to the median of 500 runs. Epidemiological parameters: $\mu=0.25, \epsilon=0.4$, $R_0=2.5$, $I_0= 100$}
    \label{panel_3d}
\end{figure*}

To show the validity of our approach beyond the two dimensions here we present the results of our analysis when we stratify the matrix along three dimensions (age, dim2 and dim3). We assume that also the third dimension has three possible groups.  Starting from a generalized contact matrix with two dimensions (Fig. \ref{panel_3d}-b)  we build a generalized contact matrix with the assumption of random mixing among the third dimension (Fig. \ref{panel_3d}-f) and assortative assumption along the third dimension (Fig. \ref{panel_3d}-e). We assume the same parameters as for the assortative regime of Fig.2 in the main text to build the 2D matrix. While for the third dimension, we assume that individuals are respectively 30\%,30\% and 40\% in the three dimensions. Additionally, we assume that 34\%,33\% and 46\%  represent the respective proportion of assortative contacts for the three groups and that 40\%,40\% and 20\%  indicate the relative activity of the three groups. Also in this case, we use these matrices to study the unfolding of a virus in the population. 
We can observe that, in the assortative mixing scenario, the spectral radius of the $G_{\mathbf{a},\mathbf{b}}$ with three dimensions (Fig. \ref{panel_3d}-e) is higher of the one of $G_{\mathbf{a},\mathbf{b}}$ with two dimensions (Fig. \ref{panel_3d}-b). This implies that, given $R0 =1$, the critical value of $\Phi$ predicted by the model with three dimensions is going to be higher than the one predicted by the model with two dimensions. 

In Fig.~\ref{panel_3d}-c we plot the attack rate (i.e., epidemic size) as a function of $R_0$ and as a function of the transmissibility parameter $\Phi$ (inset) for $1)$ a model fed only with the contact matrix $C_{ij}$ (grey circles), $2)$ a model fed with the generalized contact matrix $G_{\mathbf{a},\mathbf{b}}$ with three dimensions (purple squares). In Fig \ref{panel_3d}-d we show the prevalence of infected overtime of the three dimensions as predicted by the model fed with $G_{\mathbf{a},\mathbf{b}}$ with three dimensions for the three groups of $dim3$. Instead, in the inset we show the overall prevalence of infection predicted by the two models. We perform the same analysis for the assortative scenario and we show the results respectively in Fig~ \ref{panel_3d}-g and Fig~\ref{panel_3d}-h. All the results are in perfect agreement with the ones discussed in the main text. 

\subsection*{Real world generalized contact data}
\begin{figure*}
    \centering
    \includegraphics[width=0.83\textwidth]{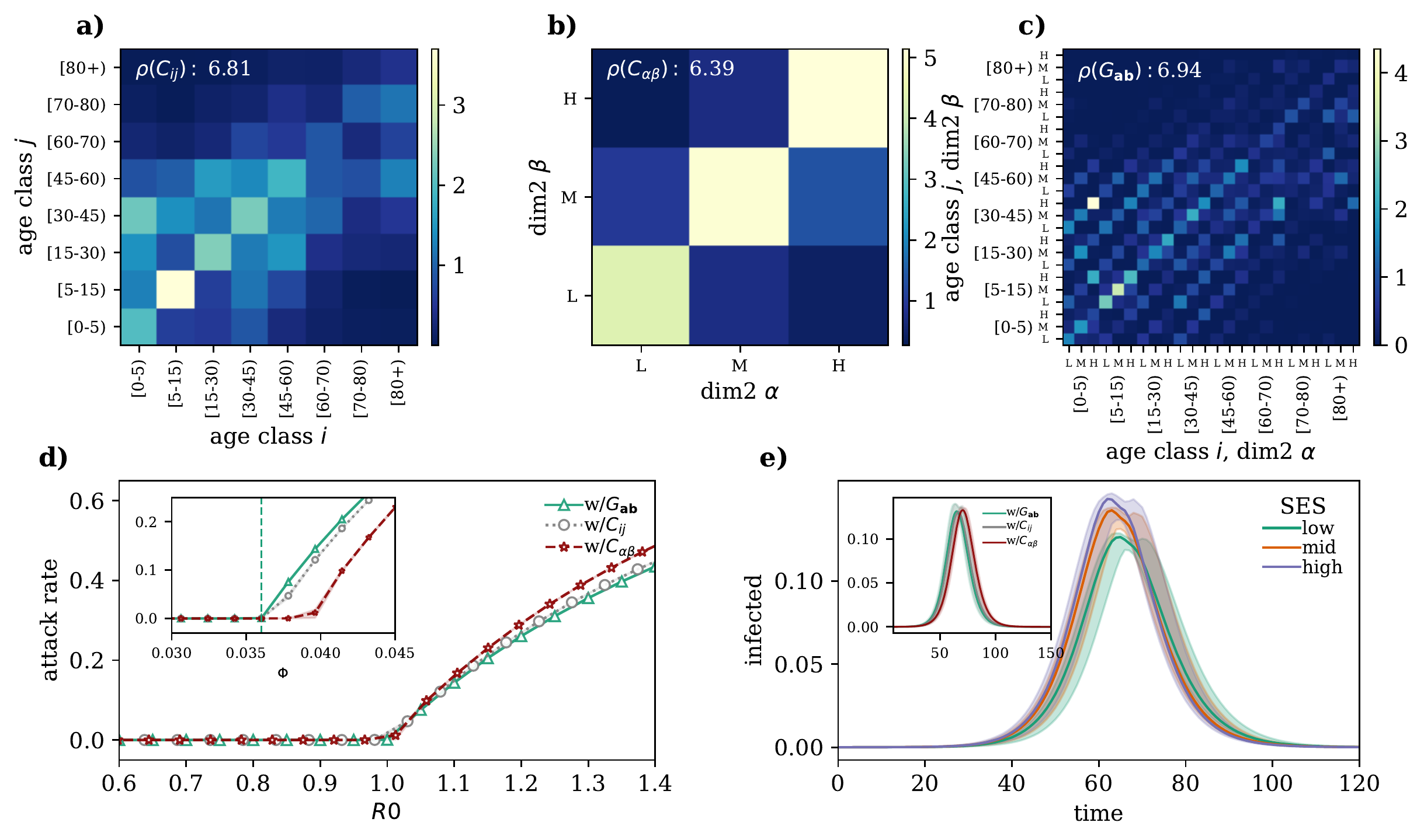}
    \caption{\textbf{Real world generalized contact matrices. }(a) Age contact matrix $C_{ij}$ ($8 \times 8$); (b) Socio-economic status (SES) contact matrix $C_{\alpha \beta}$ ($3 \times 3$) and (c) Generalized contact matrix with age and socio-economic status $G_{\mathbf{a}\mathbf{b}}$ ($24 \times 24$); (d) attack rate as a function of $R_0$ and (d) disease prevalence as a function of time. Results refer to the median of 500 runs. Epidemiological parameters: $\mu=0.25, \epsilon=0.4$, $R_0=2.7$, $I_0= 100$. The matrices have been computed using the contact diaries coming from the MASZK data collected in Hungary during April 2021.}
    \label{panel_4_12_}
\end{figure*}

\begin{figure*}
    \centering
    \includegraphics[width=0.83\textwidth]{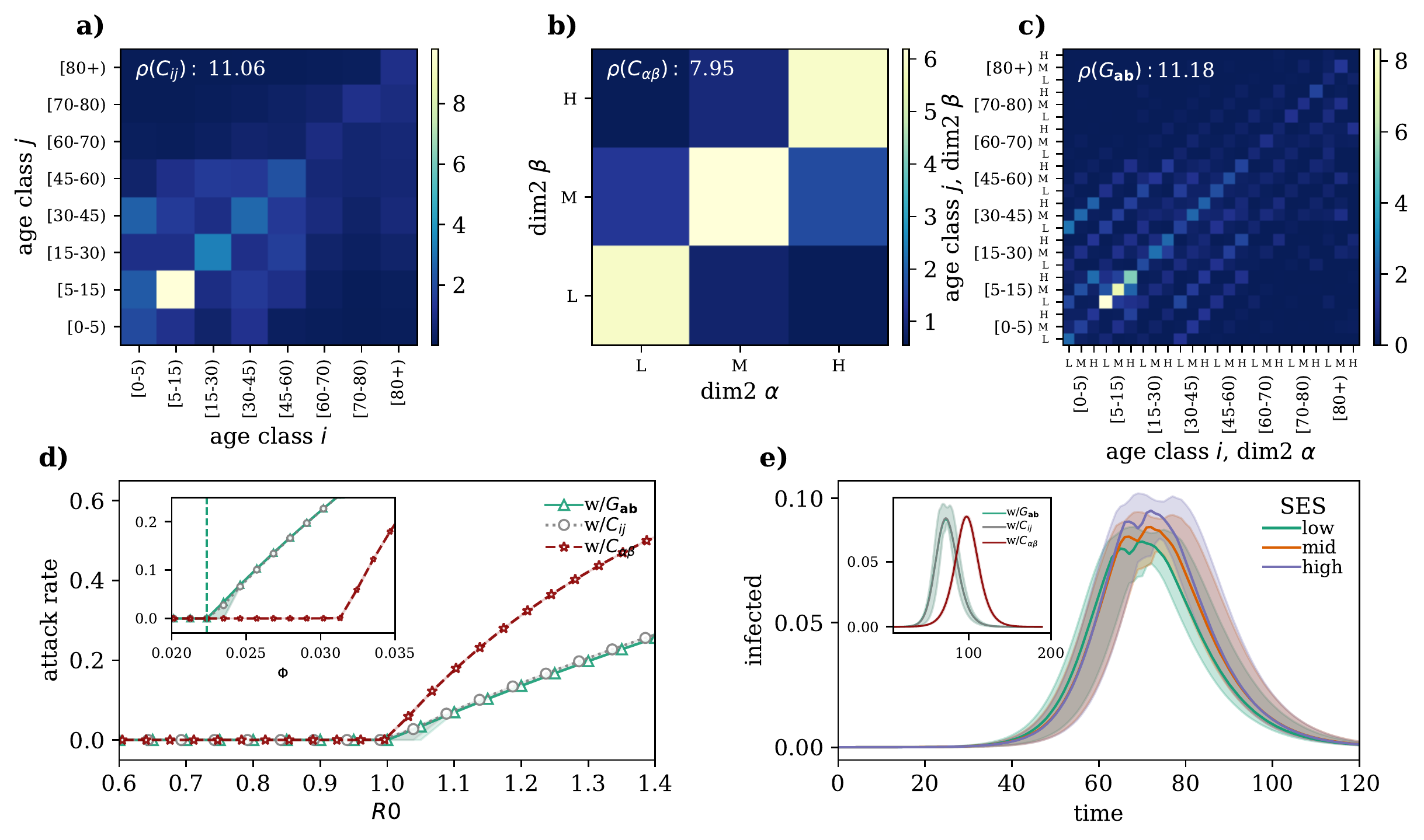}
    \caption{\textbf{Real world generalized contact matrices. }(a) Age contact matrix $C_{ij}$ ($8 \times 8$); (b) Socio-economic status (SES) contact matrix $C_{\alpha \beta}$ ($3 \times 3$) and (c) Generalized contact matrix with age and socio-economic status $G_{\mathbf{a}\mathbf{b}}$ ($24 \times 24$); (d) attack rate as a function of $R_0$ and (d) disease prevalence as a function of time. Results refer to the median of 500 runs. Epidemiological parameters: $\mu=0.25, \epsilon=0.4$, $R_0=2.7$, $I_0= 100$. The matrices have been computed using the contact diaries coming from the MASZK data collected in Hungary during November 2021.}
    \label{panel_4_19}
\end{figure*}

The MASZK questionnaire provide us with three diaries that we were able to use to generate the generalized contact matrices stratifying by age and SES of participants. Here we show the results of the same analysis shown in the main text for two additional periods: April (Fig. \ref{panel_4_12_}) and November (Fig. \ref{panel_4_19}) 2021.
While the overall results confirm the one presented in the main text, models fed with standard and generalized contact matrices lead to estimations of the attack rates, for a given $R_0$, which are closer to each other.  Nevertheless, the model featuring generalized contact matrices allows capturing heterogeneities in the incidence across SES groups which are invisible to standard approaches.

\end{document}